\documentclass[%
reprint,
superscriptaddress,
 amsmath,amssymb,
 aps,
prb,
]{revtex4-2}

\usepackage{graphicx}
\usepackage{dcolumn}
\usepackage{bm}
\usepackage{braket}
\usepackage{xcolor}
\usepackage{cancel}
\usepackage{amsmath}
\usepackage{amssymb}
\usepackage[normalem]{ulem}

\newcommand{\blue}[1]{\textcolor{black}{#1}}

\begin{document}
%

\title{Orbital-Selective Mott and Antiferromagnetic Phases in Diagonally Compressed Kagome Lattice}

\author{Jiewei Ding}
\affiliation{
Department of Physics, City University of Hong Kong, Kowloon, Hong Kong
}%
\author{Ho-Kin Tang}
\email{denghaojian@hit.edu.cn}
\affiliation{%
  School of Science, Harbin Institute of Technology, Shenzhen, 518055, China
}%
\affiliation{Shenzhen Key Laboratory of Advanced Functional Carbon Materials Research and Comprehensive Application, Shenzhen 518055, China.}
\author{Wing Chi Yu}
\email{wingcyu@cityu.edu.hk}
\affiliation{
Department of Physics, City University of Hong Kong, Kowloon, Hong Kong
}
%

\date{\today}

\begin{abstract}
We perform determinant quantum Monte Carlo simulations of the half-filled Hubbard model on a diagonally compressed kagome lattice, introducing exponential decay long–range hopping \(t(r) = t_0 \exp\bigl(-r / r_0\bigr)\) to account for the evolving bond length. By varying the lattice angle $\theta$ and the on–site interaction $U$,  double occupancy, charge compressibility, and spin–spin correlation functions of the whole system and each sub-lattice are measured. We find that geometric compression induces a clear sublattice differentiation: for $\theta\gtrsim52^\circ$, the A sublattice establishes  long-range hoppings, which in turn suppresses the metallic behavior of the $B/C$ sublattice and drives a selective Mott transition; for $\theta\lesssim52^\circ$, the $B$–$C$ chains develop \blue{long-range antiferromagnetic correlations within the finite-size simulations}, which in turn suppresses the metallic behavior of the $A$ sublattice and drives a selective Mott transition. \blue{The critical interaction $U^c_A$ for the $A$ sites decreases sharply near the onset of $B$--$C$ antiferromagnetic correlations}, while $U^c_{B/C}$ increases. These competing orders give rise to an orbital–selective Mott phase and a rich $U$–$\theta$ phase diagram featuring paramagnetic–metal, paramagnetic–Mott, antiferromagnetic–metal, and antiferromagnetic–Mott states. Our results highlight the complex interplay between lattice geometry, magnetic frustration, and strong correlations in frustrated two–dimensional systems.
\end{abstract}

\maketitle


\section{Introduction}
The frustration arising from geometric structure of the lattice, where not all spin interactions can simultaneously satisfy their lowest-energy configurations, plays a central role in exotic states in condensed matter systems. In two-dimensional lattices, such as triangular and kagome lattices, geometric frustration suppresses conventional N\'eel order, resulting in ground-state degeneracy and novel quantum phases. For instance, the $S=\tfrac{1}{2}$ kagome lattice avoids long-range magnetic order due to corner-sharing triangles, instead realizing a quantum spin-liquid state characterized by persistent spin fluctuations and an absence of symmetry breaking even at the lowest temperatures~\cite{balents2010spin,zhou2017quantum,helton2007spin,han2012fractionalized}. Similarly, triangular lattice exhibits a spin-liquid phase resulting from the interplay of strong electronic correlations and geometric frustration, preventing both magnetic ordering and metallic conductivity~\cite{shimizu2003spin}. In weakly correlated electronic systems, kagome metals such as Mn$_3$Sn exhibit a non-collinear antiferromagnetic structure, leading to a significant anomalous Hall effect in zero magnetic field ~\cite{nakatsuji2015large,ikhlas2017large}. Meanwhile, ferromagnetic kagome metals like Fe$_3$Sn$_2$ display large gaps at Dirac points, creating massive Dirac fermions. It also gives rise to pronounced anomalous Hall responses and topological electronic states~\cite{ye2018massive}. Another family of kagome metals, AV$_3$Sb$_5$ (A = K, Rb, Cs) and CsCr$_3$Sb$_5$, has also been found to host charge density wave order, non-trivial topology, and superconductivity \cite{Ortiz2019-el,Neupert2022-tu,Zhang2024-lb,Wang2025-hq,Liu2024-vq}. Furthermore, it has been recently predicted from density functional theory calculations that CsCr$_3$Sb$_5$ could exhibit an altermagnetic state under pressure \cite{Xu2025-zk}. These examples underscore how geometric frustration fundamentally affect the properties of two-dimensional materials.~\\

\blue{Electron-electron interactions constitute another crucial factor influencing the physical properties of lattice systems. At half filling, when the Coulomb repulsion is sufficiently strong, electrons can become localized, giving rise to Mott insulating states that are often accompanied by magnetic correlations~\cite{imada1998metal,lee2006doping}. The half-filled regime therefore provides a natural starting point for studying strong correlation effects. Experimentally, correlated superconductors such as cuprates and iron-based oxypnictides exhibit superconductivity when carriers are doped into or removed from their parent compounds~\cite{orenstein2000advances,gull2013superconductivity,keimer2015quantum,zou2024evolution,chen2008superconductivity,bernardini2018iron}, highlighting the broader relevance of understanding correlated electronic states near half filling.} More generally, proximity to a Mott transition can induce diverse phases; for example, in V$_2$O$_3$, pressure or doping drives transitions between metallic and insulating phases~\cite{imada1998metal}. On frustrated lattices, Hubbard models similarly show finite-temperature Mott transitions, exemplified by the anisotropic triangular-lattice Hubbard model with a critical interaction $U^c/t \approx 9.1$ and a critical temperature $T/t \approx 0.3$~\cite{ohashi2008finite}. Consequently, understanding Mott transitions is essential for clarifying electronic phase diagrams in geometrically frustrated lattices.~\\

\blue{The half-filled kagome lattice Hubbard model illustrates} intense competition among kinetic energy, Coulomb repulsion, and geometric frustration. Prior studies employing variational cluster approximation, dynamical mean-field theory, variational Monte Carlo, and determinant quantum Monte Carlo (DQMC) simulations yielded varying Mott transition critical values, typically in the range of $U^c/t\in[0,11]$~\cite{yamada2011mott,higa2016bond,ohashi2006mott,kuratani2007mott,guertler2014kagome}. Recent density-matrix renormalization group calculations reveal a continuous metal-insulator transition at $U^c/t \approx 5.4$, subsequently evolving into valence-bond crystal and quantum spin-liquid phases, eventually linking to the Heisenberg limit at strong coupling~\cite{sun2021metal}. Latest DQMC simulations suggest a paramagnetic-metal to Mott-insulator transition occurs at $U^c/t \approx 6.5$ without developing magnetic order in the insulating phase~\cite{medeiros2023thermodynamic,kaufmann2021correlations}.~\\

In the kagome lattice under half-filling, adjusting geometric frustration elucidates how frustration competes with electron-electron interactions, affecting phase transitions. One approach to tune the degree of frustration in the kagome lattice is to introduce an additional site at the center of each hexagon, with hopping amplitude $t'$ connecting it to the surrounding sites. As $t'$ increases from 0 to $t$, the system interpolates from the kagome lattice to the triangular lattice~\cite{bulut2005magnetic}. A more common method is to introduce bond anisotropy, by setting one set of bonds to $t'$ while keeping the other two sets at $t$, thereby continuously interpolating between the fully frustrated kagome lattice ($t'/t=1$) and the non-frustrated Lieb lattice ($t'/t\to0$)~\cite{lima2023effects,lim2020dirac,jiang2019topological,yamada2011mott,jiang2019topologicalErratum}. Recent DQMC studies observed this stretched lattice’s phase diagram, observing monotonic increase of $U^c$ with enhanced frustration and disappearance of ferromagnetic order near $t'/t\approx0.55$~\cite{lima2023magnetism}. Though challenging to realize in actual materials, kagome lattices with adjustable interactions are experimentally achievable in cold atom setups~\cite{jo2012ultracold,chern2014dynamically,leung2020interaction}, offering a testable platform. Thus, continued investigation of systems with tunable frustration remains significant for guiding experiments.~\\

Previous studies of frustration tuning have primarily focused on the kagome–Lieb model, while the regime of extreme compression from kagome toward quasi-one-dimensional chains has been overlooked. We propose an alternative method for tuning frustration, namely, diagonal compression, in which the angle between the primitive vectors 
is reduced from $60^\circ$ to a smaller angle $\theta$, thereby geometrically compressing the lattice. On the one hand, this geometric deformation shortens the $BC$ bond length [Fig. \ref{fig:fig1} (a, c)], thereby reducing triangular frustration; on the other hand, by assuming an exponential decay of the hopping amplitude with distance, next-nearest-neighbor hopping can be naturally incorporated into the Hubbard model. Compared to models with simple bond anisotropy, this compressed kagome lattice more realistically captures the impact of geometric deformation on the magnetic and electronic properties, providing a new perspective for studying the interplay among geometry, long-range hopping, and strong correlations.~\\

Using DQMC simulations of the half-filled
compressed kagome Hubbard model, we find that, under weak compression, increasing $U$ drives the system from a paramagnetic metal to a paramagnetic Mott insulator. Under strong compression, increasing $U$ drives transitions from a paramagnetic metal to an antiferromagnetic metal and subsequently to an antiferromagnetic Mott insulator. Sublattice analysis reveals that geometric compression breaks the equivalence among the three sublattices and leads to orbital-selective Mott behavior: for strong compression, electrons on sublattice $A$ are more easily localized and insulating, while sublattices $B$ and $C$ remain more metallic; for weak compression, the situation is reversed.~\\

The paper is organized as follows: Sec.~\ref{sec:model_method} introduces the model and methodology; Sec.~\ref{sec:simulation_results} presents simulation results including conductivity, magnetization, and phase diagrams in the $U$–$\theta$ plane; Sec.~\ref{sec:conclusion} concludes the paper.~\\

\section{The model and the observables}
\label{sec:model_method}

\begin{figure} [t!]
\includegraphics[width=0.99\linewidth]{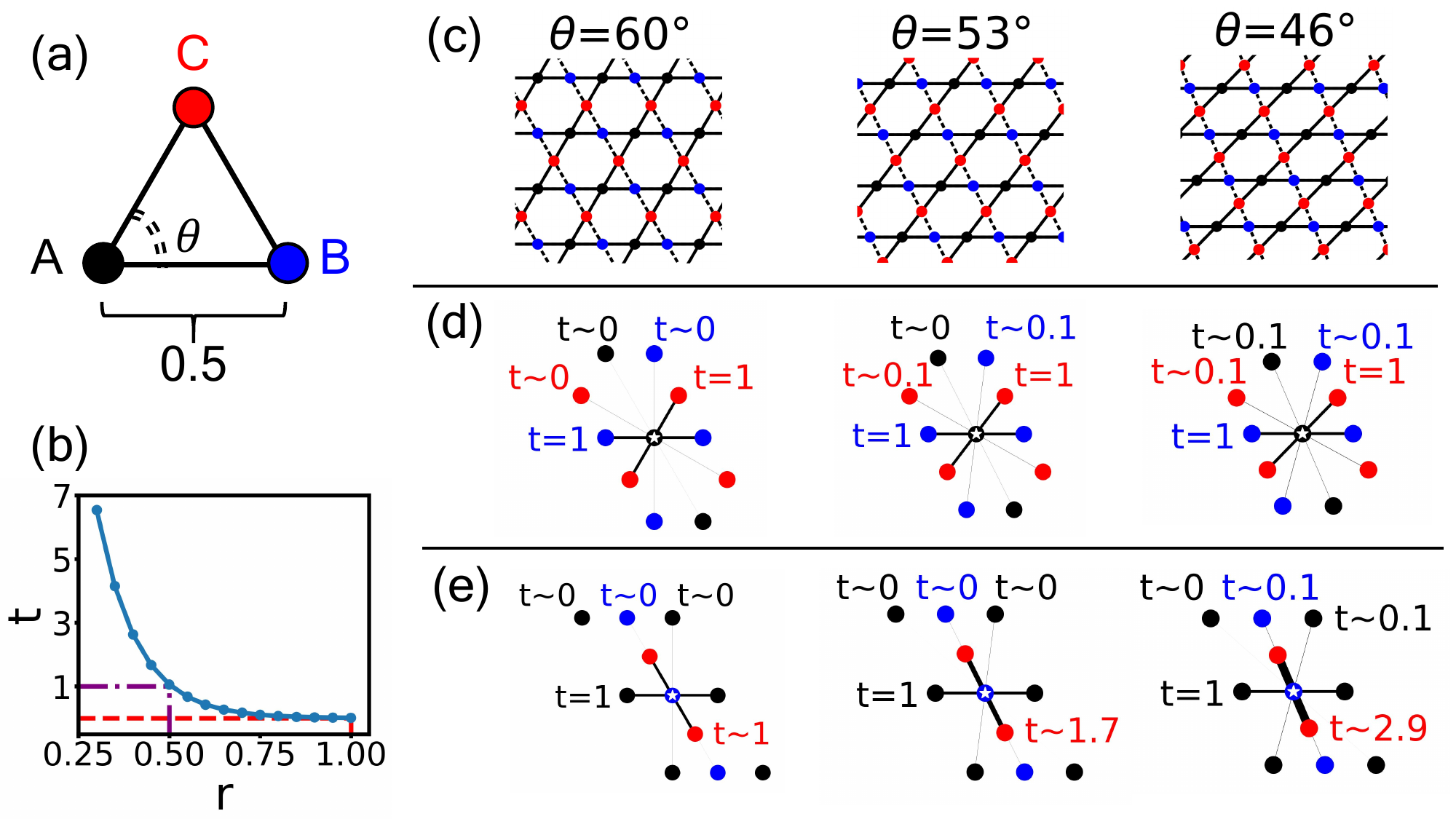}
\caption{(a) Schematic of the kagome lattice before and after compression. Solid lines indicate fixed $AB$ and $AC$ bond lengths; the dashed line shows the $BC$ bond shortening as the angle $\theta$ decreases. (b) Exponential decay of hopping strength $t(r)$ with distance $r$. (c) Evolution of the lattice geometry with $\theta$. (d) and (e): Hopping amplitudes between sublattices $A$ and $B$ and their 10 nearest sites, with line thickness proportional to $t$.}
\label{fig:fig1}
\end{figure}

We investigate the Hubbard model with repulsive on-site interactions on a diagonally compressed kagome lattice. As shown in Fig.~\ref{fig:fig1}(a), the original kagome lattice consists of three sublattices $A$, $B$, and $C$ in an unit cell, with the spacing between adjacent sites within the same unit cell set to $0.5$. Let $\mathbf{r}_A$ denote the position vector of a site on sublattice $A$; the positions of sublattices $B$ and $C$ are then given by $\mathbf{r}_B = \mathbf{r}_A + 0.5\,\hat{\mathbf{x}}$ and $\mathbf{r}_C = \mathbf{r}_A + 0.25\,\hat{\mathbf{x}} + \frac{\sqrt{3}}{4}\,\hat{\mathbf{y}}$, respectively. At this point, the angle between the $AB$ and $AC$ bonds is $\theta=60^\circ$. Under diagonal compression, the $AB$ and $AC$ bond lengths are held fixed, while $\theta$ is gradually reduced, resulting in a continuous shortening of the $BC$ bond. As the $BC$ bond length decreases, electron hopping between distant sites becomes significant; thus, long-range hopping is introduced into the Hubbard model, parameterized as $t(r) = t_0 \exp\bigl(-r / r_0\bigr)$, where $t_0 = 100$ and $r_0 = 0.11$, as illustrated in Fig.~\ref{fig:fig1}(b). When $r=0.5$, $t \approx 1.0$; when $r=1.0$, $t \approx 0.011$, which recovers the nearest-neighbor hopping of the kagome lattice ($\theta=60^\circ$). \blue{Equivalently, specifying \(t(0.5)\) and \(t(1.0)\) fixes the two parameters in the exponential form up to the rounding used here.}~\\

As $\theta$ decreases, the solid lines in Fig.~\ref{fig:fig1}(c) indicate the fixed $AB$ and $AC$ bond lengths, while the dashed line shows the shortening of the $BC$ bond with decreasing $\theta$. Figures~\ref{fig:fig1}(d) and (e) illustrate the hopping amplitudes $t$ from sublattices $A$ and $B$ to their ten nearest sites, with line thickness proportional to $t$. As $\theta$ decreases from $60^\circ$ to $46^\circ$, the nearest-neighbor hopping amplitudes from $A$ to $B$ and $C$ remain unchanged, while the next-nearest-neighbor hopping increases slightly from $0$ to about $0.1$; similarly, the nearest-neighbor hopping from $B$ to $A$ remains unchanged, while the hopping amplitude $B$ to $C$ increases from $1$ to approximately $2.9$, and the next-nearest-neighbor hopping also increases slightly from $0$ to about $0.1$. \blue{In the numerical simulations, for each site \(i\) at $\mathbf{r_i}$, we retain hopping processes only to its ten nearest neighbors, with the corresponding amplitudes determined by \(t(|\mathbf r_i-\mathbf r_j|)\). No additional amplitude cutoff is imposed within these ten sites.}~\\

The Hubbard Hamiltonian with long-range hopping is then given by
\begin{equation}\label{eq:Hubbard}
\begin{aligned}
H &= - \sum_{i\neq j,\sigma} t\bigl(|\mathbf{r}_i - \mathbf{r}_j|\bigr)\, 
\bigl(c^{\dagger}_{i\sigma} c_{j\sigma} + \mathrm{H.c.}\bigr)
- \mu \sum_{i,\sigma} n_{i\sigma} \\
&\quad + U \sum_i \left(n_{i\uparrow}-\frac{1}{2}\right)\left(n_{i\downarrow}-\frac{1}{2}\right),
\end{aligned}
\end{equation}
where $t(|\mathbf{r}_i - \mathbf{r}_j|)$ is the exponentially decaying hopping amplitude we mentioned above, $\mu$ is the chemical potential controlling the particle number, and $U$ is the on-site Coulomb repulsion. The operators $c^{\dagger}_{i\sigma}$ ($c_{i\sigma}$) create (annihilate) a fermion with spin $\sigma$ at site $i$, with $n_{i\sigma}=c^{\dagger}_{i\sigma}c_{i\sigma}$ and $\sigma = \uparrow,\downarrow$ denoting spin-up and spin-down electrons, respectively.~\\

To investigate the thermodynamic properties at finite temperature, we employ the DQMC method. DQMC is a widely used numerical technique, which utilizes the Trotter–Suzuki decomposition and Hubbard–Stratonovich (HS) transformation to rewrite the original Hamiltonian (with a Hilbert space dimension of $4^N$, where $N$ is the number of sites) into a form that evolves only $N$ basis states in imaginary time~\cite{blankenbecler1981monte,santos2003introduction,uo2007lecture,qin2016benchmark}.~\\

Specifically, the partition function $Z$ of the original Hamiltonian is first decomposed into a product of exponential operators using the Trotter–Suzuki approximation,
\begin{equation}
Z = \mathrm{Tr}\left(e^{-\beta H}\right)
  = \mathrm{Tr}\left[\left(e^{-\delta\tau (H_K + H_U)}\right)^{L_t}\right],
\end{equation}
where $\beta = 1/T$ is the inverse temperature (with Boltzmann constant $k_B=1$), $\delta\tau$ is the imaginary-time interval, $H_K$ and $H_U$ correspond to the hopping plus chemical potential term and the on-site interaction term in the Hamiltonian, respectively, and $L_t$ is the number of time slices such that $\beta = \delta\tau \times L_t$. This decomposition introduces an error of order $O(\delta\tau^2)$, which is negligible for $\delta\tau=0.1$ as used in this study.

Since $\delta\tau$ is small, $H_K$ and $H_U$ can be approximately considered as commuting, further separating the partition function as
\begin{equation}
Z \approx \mathrm{Tr}\left[\left(e^{-\delta\tau H_K} e^{-\delta\tau H_U}\right)^{{L_t}}\right].
\end{equation}

Here, $H_K$ describes free fermions and can be represented in the single-particle basis, while $H_U$ contains the interaction term $n_{i,\uparrow} n_{i,\downarrow}$. We decouple $H_U$ using the Hubbard–Stratonovich transformation, introducing an auxiliary field that couples to the local spin,
\begin{equation}
e^{-U\delta\tau\,(n_{i,\uparrow}-\frac{1}{2})(n_{i,\downarrow}-\frac{1}{2})}
= \frac{1}{2}e^{-U\delta\tau/4}
  \sum_{h_i=\pm1}e^{\nu h_i (n_{i,\uparrow} - n_{i,\downarrow})},
\end{equation}
where $\cosh(\nu) = e^{U\delta\tau/2}$. After this transformation, $n_{i,\uparrow}$ and $n_{i,\downarrow}$ are decoupled, at the expense that both spins on each site $i$ couple to an auxiliary field $h_i = \pm1$. As there are $L_t$ factors of $e^{-\delta\tau H_U}$ in the partition function, we define an $N \times L_t$ auxiliary field matrix $h$, with element $h_{i,\ell}$ representing the auxiliary field at site $i$ and time slice $\ell$. The transformed partition function takes the form
\begin{equation}
Z \propto \sum_{h} Z_h,
\end{equation}
with
\begin{equation}
Z_h = \det M_{\uparrow}(h)\,\det M_{\downarrow}(h),
\end{equation}
and
\begin{equation}
M_{\sigma}(h) = I + B_{L_t,\sigma}(h) B_{L_t-1,\sigma}(h) \cdots B_{1,\sigma}(h),
\end{equation}
where $I$ is the identity matrix and $B_{\ell,\sigma}(h)$ denotes the matrix associated with the Trotter–Suzuki decomposition of the Hamiltonian following the Hubbard–Stratonovich transformation. In this way, the quantum system is mapped to a semiclassical system amenable to classical Monte Carlo sampling over the auxiliary field configurations $h$. \blue{The normalized statistical weight of each configuration is $P(h)=Z_h/Z$}, and the relevant observables can be computed using the corresponding Green's function $G_{\sigma}(h) = [ M_{\sigma}(h)]^{-1}$. However, unlike classical systems where the Boltzmann weight $P = e^{-E\beta}$ is always non-negative, the weight of each HS-field $Z_h$ in DQMC can be negative. To ensure positive sampling weights, the probability is taken as $|Z_h|$, and the sign $s(h) = \mathrm{sign}(Z_h)$ is included as a correction in the observable averages~\cite{santos2003introduction,iglovikov2015geometry,pan2022sign},
\begin{equation}
\langle O \rangle
= \frac{\sum_{h}s(h)\,|Z_h|\,O(h)}
       {\sum_{h}s(h)\,|Z_h|}
= \frac{\langle sO\rangle}{\langle s\rangle}.
\end{equation}
~\\

\begin{figure}[t!]
\centering
\includegraphics[width=0.75\linewidth]{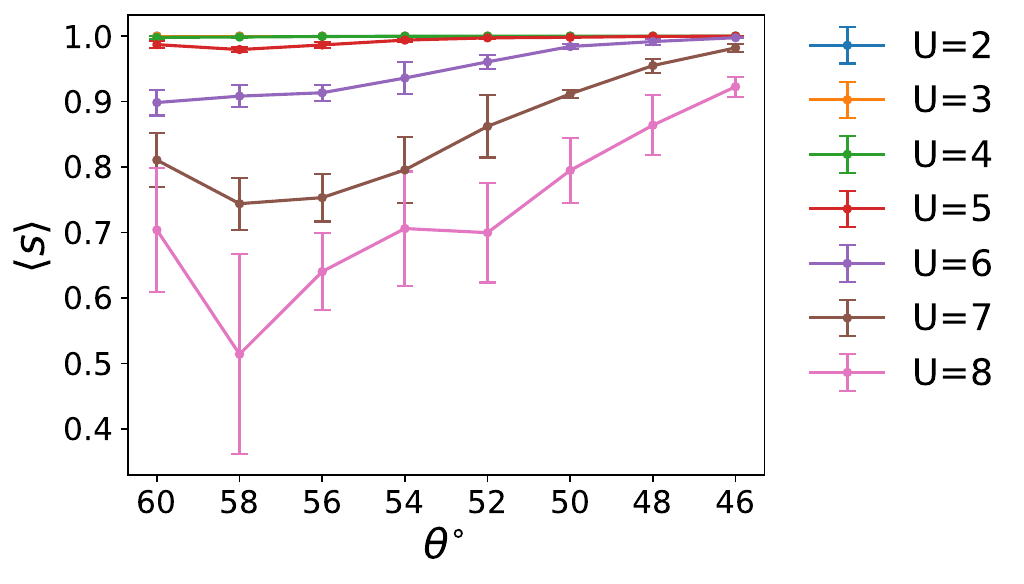}
\caption{\blue{
Average sign $\langle s\rangle$ at the lowest temperature $T=0.5$
as a function of the compression angle $\theta$ along the
half-filling line. Different curves correspond to different values of
$U$. For each $(U,\theta)$, the chemical potential is fixed by the half-filling condition $n=1$. Error bars denote the standard
error estimated from independent Monte Carlo simulations.}
}
\label{fig:sign}
\end{figure}

\blue{The severity of the sign problem is characterized by the average sign $\langle s\rangle$. Since the main results of this work are obtained at half-filling, we evaluate $\langle s\rangle$ along the corresponding half-filled line, where the chemical potential is chosen such that the average density $n=\blue{\frac{1}{N}\sum_{i,\sigma}\langle n_{i\sigma}\rangle}=1$ for each set of $(T,U,\theta)$. As shown in Fig.~\ref{fig:sign}, at the lowest temperature used in this work, $T=0.5$, the average sign remains reasonably large for the parameter range studied here. The compression does not worsen the sign problem; instead, $\langle s\rangle$ generally increases as $\theta$ is reduced. Thus, the lowest temperature $T=0.5$ is mainly limited by the computational cost of scanning a broad parameter space with multiple independent Markov chains, rather than by a severe sign problem.}~\\

To probe the critical behavior of the system, we measure several physical quantities, including the double occupancy $D$, electronic compressibility $\kappa$, the the spin-spin correlation function $G(r)$, and the local magnetization $M$ in the $z$ direction. The definitions of double occupancy and electronic compressibility are given by 
\begin{equation}
D=\frac{1}{N}\sum_{i=1}^{N} \langle n_{i,\uparrow} n_{i,\downarrow}\rangle,
\end{equation}
and
\begin{equation}
\kappa = \frac{1}{\langle n \rangle^{2}}\frac{\partial \langle n \rangle}{\partial \mu},
\end{equation}
where $n$ is the average particle number per site. The double occupancy $D$ measures a lattice site is simultaneously occupied by both spin-up and spin-down electrons, for half-filling, $D \in [0,\,0.25]$. In a half-filled system, if $\kappa$ approaches zero, it indicates that the average particle number per site does not increase with increasing $\mu$. Each site would accommodate only one electron, and double occupancy is suppressed; in this case, the system enters the Mott insulating phase~\cite{medeiros2023thermodynamic}. That is, electrons become localized due to the forbidden of double occupancy, thereby suppressing conductivity.~\\

\blue{For sublattice-resolved quantities, we do not introduce independent sublattice-dependent chemical potentials. Instead, we define the sublattice-resolved charge response as
\begin{equation}
\kappa_{\alpha} = \frac{\partial \langle n_{\alpha}\rangle}{\partial \mu},
\end{equation}
where $\alpha=A,B/C$ and the same global chemical potential $\mu$ is adopted for the whole system. Numerically, $\kappa$ and $\kappa_{\alpha}$ are extracted from the slopes of $\langle n\rangle $ and $\langle n_{\alpha}\rangle$, respectively, with respect to the same global chemical potential $\mu$.}~\\

The spin operators at site $i$ are defined as follows:
\begin{equation}
S_i^z = n_{i,\uparrow} - n_{i,\downarrow},
\end{equation}
and
\begin{equation}
S_i^x = c_{i\uparrow}^\dagger\,c_{i\downarrow} + c_{i\downarrow}^\dagger\,c_{i\uparrow}.
\end{equation}
The spin–spin correlation functions in the $z$ and $x$ directions, $G^{z}(r)$ and $G^{x}(r)$, are defined via Wick’s theorem as~\cite{wick1950evaluation}:
\begin{equation}
\begin{split}
G^{z}(r) 
&= \langle S^z_i S^z_{i+r}\rangle \\
&= \bigl(\langle c^\dagger_{i\uparrow}c_{i\uparrow}\rangle
         - \langle c^\dagger_{i\downarrow}c_{i\downarrow}\rangle\bigr)\,
  \bigl(\langle c^\dagger_{i+r,\uparrow}c_{i+r,\uparrow}\rangle
         - \langle c^\dagger_{i+r,\downarrow}c_{i+r,\downarrow}\rangle\bigr)\\
&\quad
  + \langle c^\dagger_{i\uparrow}c_{i+r,\uparrow}\rangle\,
    \langle c_{i+r,\uparrow}c^\dagger_{i,\uparrow}\rangle
  + \langle c^\dagger_{i\downarrow}c_{i+r,\downarrow}\rangle\,
    \langle c_{i+r,\downarrow}c^\dagger_{i,\downarrow}\rangle,
\end{split}
\label{eq:Gz_r}
\end{equation}

\begin{equation}
\begin{split}
G^{x}(r)
&= \langle S^x_i S^x_{i+r}\rangle \\
&= \langle c^{\dagger}_{i\downarrow}c_{i+r,\downarrow}\rangle\,
    \langle c_{i+r,\uparrow}c^{\dagger}_{i\uparrow}\rangle
  + \langle c^{\dagger}_{i\uparrow}c_{i+r,\uparrow}\rangle\,
    \langle c_{i+r,\downarrow}c^{\dagger}_{i\downarrow}\rangle,
\end{split}
\label{eq:Gx_r}
\end{equation}
where $r$ denotes the distance between two sites. By evaluating $G(r)$, one can determine the presence of long-range magnetic correlations and thus infer the global magnetic state of the system~\cite{bulut2005magnetic,wen2022superconducting,xiao2023temperature,varney2009quantum,sandvik2010computational}. \blue{The local magnetic moment in the $z$ direction is given by the on-site correlation function, $M=G^z(0)=\langle (S_i^z)^2\rangle$. Since \(S_i^z=n_{i,\uparrow}-n_{i,\downarrow}\), one obtains $M=\langle n_{i,\uparrow}+n_{i,\downarrow}-2n_{i,\uparrow}n_{i,\downarrow}\rangle=\rho-2D$, where \(\rho=\langle n_{i,\uparrow}+n_{i,\downarrow}\rangle\) is the local density and \(D=\langle n_{i,\uparrow}n_{i,\downarrow}\rangle\) is the local double occupancy.}~\\

\blue{To characterize the magnetic correlations in momentum space, we calculate the path-resolved static spin structure factor along the BC direction as the Fourier transform of the equal-time spin correlation function \cite{lima2023magnetism}, 
\begin{equation} 
S^{x}_{BC}(q) = \sum_{r=0}^{N_{BC}-1} e^{-iqr}G^{x}_{BC}(r), \label{eq:structure_factor_BC} 
\end{equation} 
where $q=2\pi n/N_{BC}$, with $n=0,1,\ldots,N_{BC}-1$, and $N_{BC}=2L$ denoting the number of distinct sites along the periodic BC path. The duplicated endpoint of the real-space path is excluded from the Fourier transform. We further quantify the sharpness of the antiferromagnetic peak at $q=\pi$ using the symmetrized correlation ratio \cite{oliveira2023magnetic}, 
\begin{equation} 
R_{\pi} = 1- \frac{ S^{x}_{BC}(\pi-\delta q) + S^{x}_{BC}(\pi+\delta q) }{ 2S^{x}_{BC}(\pi) }, \label{eq:correlation_ratio} 
\end{equation} 
where $\delta q=\frac{2\pi}{N_{BC}}$. Equation~(\ref{eq:correlation_ratio}) is the inversion-symmetrized form of the conventional correlation ratio $1-S(\mathbf{Q}+\delta\mathbf{q})/S(\mathbf{Q})$. The ratio is evaluated separately for each independent Monte Carlo run.}~\\

To ensure feasible computational cost and timescales, we set the linear system size to $L=6$, resulting in a total number of sites $N=3L^2=108$. For each set of parameters, we perform 10 independent DQMC simulations with different random initial $h$. Each simulation undergoes $5\times10^3$ warm-up sweeps to equilibrate the Markov chain, followed by $5\times10^4$ measurement sweeps. \blue{Here, a sweep consists of both a forward and a backward pass over the imaginary-time slices, corresponding to \(2NL_t\) local Monte Carlo updates per sweep. This forward-backward procedure is only an algorithmic update scheme for sampling the Hubbard-Stratonovich fields more symmetrically over imaginary time and does not modify the Hamiltonian or the measured observables.~\\}~\\

\section{Results}
\label{sec:simulation_results}

\begin{figure}[t!]
\centering
\includegraphics[width=0.99\linewidth]{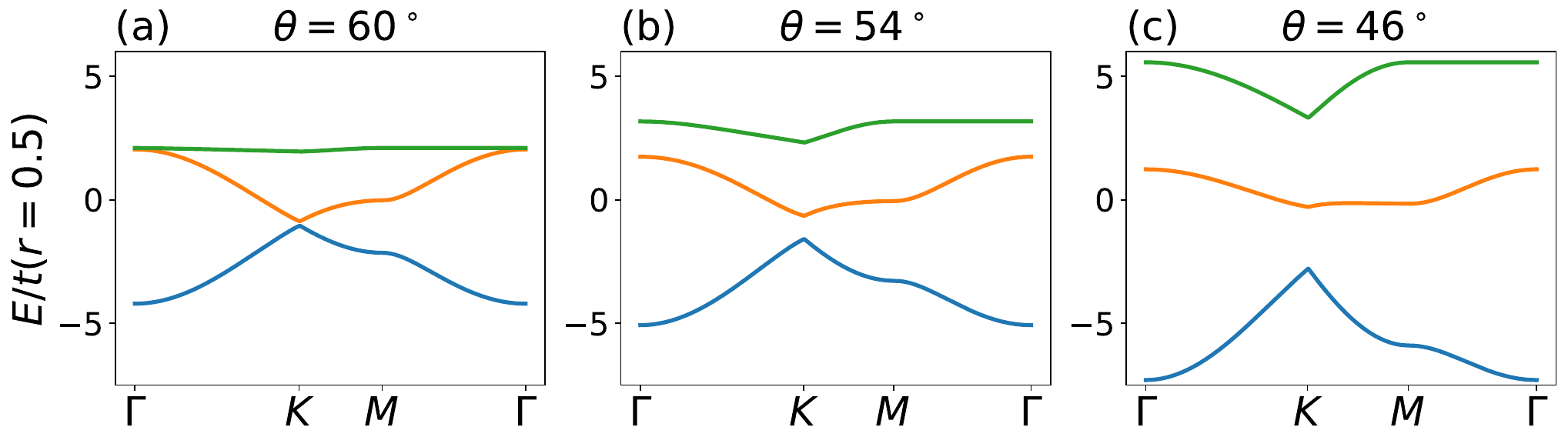}
\caption{\blue{Band structures of the diagonally compressed kagome lattice in the non-interacting limit for $\theta=60^\circ$, $54^\circ$, and $46^\circ$, obtained by diagonalizing the $3\times3$ Bloch Hamiltonian with the same distance-dependent hopping used in the DQMC simulations at $U=0$ and $\mu=0$. Energies are measured in units of the nearest-neighbor hopping $t$ with $r=0.5$ of the undistorted lattice. As $\theta$ decreases, the enhanced $BC$ hopping renders the kagome flat band dispersive and increases the overall bandwidth.}}
\label{fig:fig2_U=0}
\end{figure}

\blue{We first examine the noninteracting band structure of the compressed lattice. This provides a single-particle reference for understanding how the diagonal compression modifies the kinetic energy scale and breaks the equivalence among the three sublattices. Figure~\ref{fig:fig2_U=0} shows the band structures obtained by diagonalizing the $3\times3$ Bloch Hamiltonian constructed from the same distance-dependent hopping amplitudes used in the DQMC simulations.}~\\

\blue{At $\theta=60^\circ$, the spectrum reproduces the characteristic features of a kagome system, namely a nearly flat band and a Dirac crossing point at the two dispersive bands. The upper band is not perfectly flat because the exponential hopping form retains small but finite long-range hopping amplitudes. Upon reducing $\theta$, the $BC$ bond is shortened and the corresponding hopping is enhanced. Consequently, the flat band becomes dispersive and the total bandwidth increases. This single-particle spectrum reconstruction is most pronounced at $\theta=46^\circ$, where the large $BC$ hopping produces strong sublattice differentiation. The interacting DQMC results discussed below show how this  anisotropy is further amplified by the Hubbard interaction.}~\\

\begin{figure} [t!]
\includegraphics[width=0.323\linewidth]{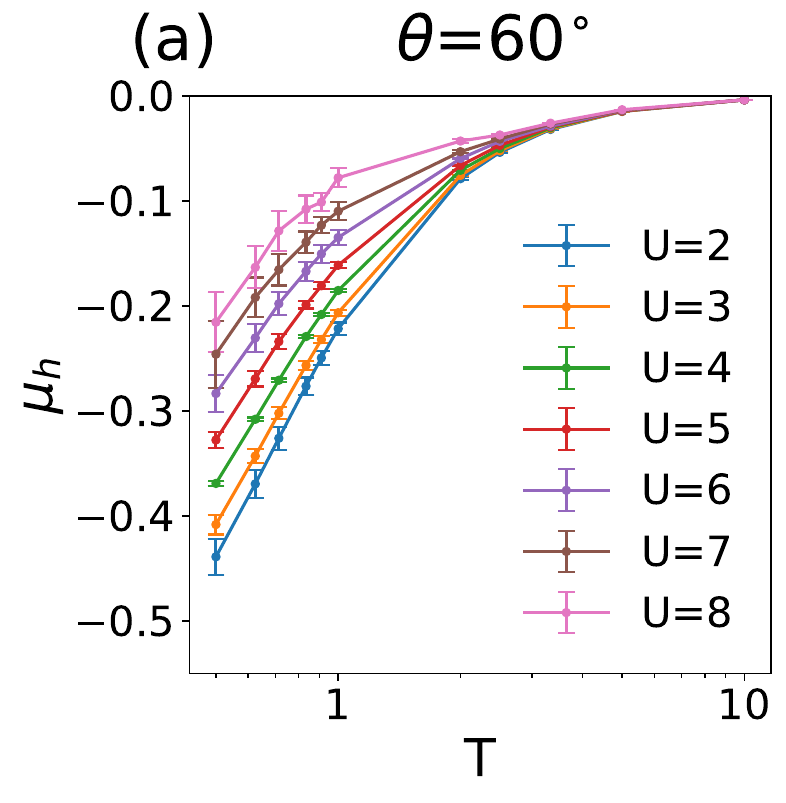}
\includegraphics[width=0.323\linewidth]{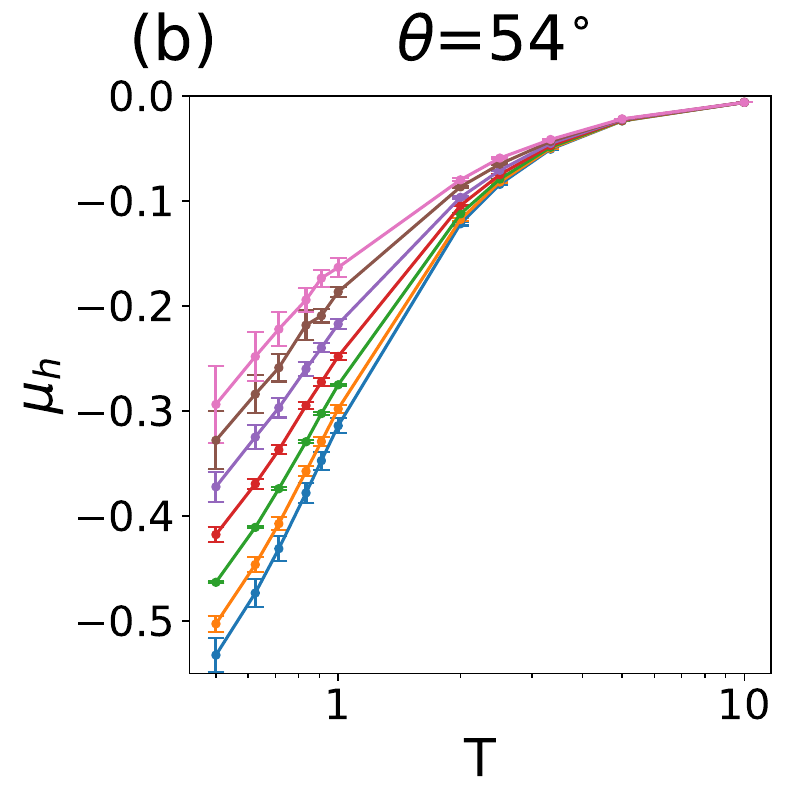}
\includegraphics[width=0.323\linewidth]{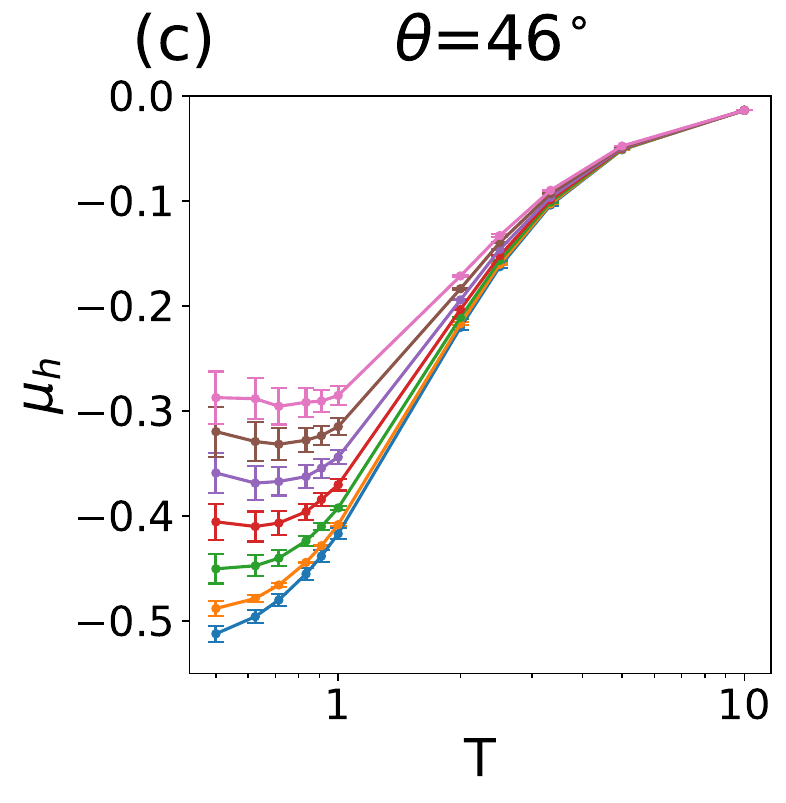}
\caption{Temperature dependence of the global chemical potential $\mu_h$ required to impose half filling ($n=1$) for various interaction strengths $U$. Panels (a-c) show $\mu_h(T)$ for different lattice angles $\theta$. Error bars indicate the standard error estimated from ten independent Monte Carlo simulations.}
\label{fig:fig2}
\end{figure}

\blue{We next determine the global chemical potential used to impose the half-filled condition in the interacting DQMC simulations. Because the compressed kagome lattice with distance-dependent hopping does not in general possess an exact particle-hole symmetry, the half-filled point cannot be fixed a priori from symmetry. We therefore determine the global chemical potential numerically for each set of parameters $(U,T,\theta)$. Specifically, for each set of $(U,T,\theta)$, we simulate a series of global chemical potentials that bracket the half-filled condition $n=1$. In the vicinity of the crossing between the calculated density curve $n(\mu)$ and $n=1$, the chemical potentials are sampled with a spacing $\Delta\mu=0.05$. The corresponding half-filled chemical potential $\mu_h$ is then determined by linear interpolation between the two neighboring data points that enclose $n=1$.  All sublattice-resolved observables discussed below are evaluated at this same global chemical potential $\mu_h$; no sublattice-dependent chemical potential is introduced. Figure~\ref{fig:fig2} displays the resulting $\mu_h$ as a function of temperature for different interaction strengths $U$ and lattice angles $\theta$. Near intermediate-to-low temperatures, reducing the angle from $\theta=60^\circ$ to $54^\circ$ lowers $\mu_h$ and enhances its temperature dependence. Further compression to $\theta=46^\circ$ weakens this temperature dependence, leading to a plateau-like behavior at large $U$.}~\\

\begin{figure}[t!]
\centering
\includegraphics[width=0.40\linewidth]{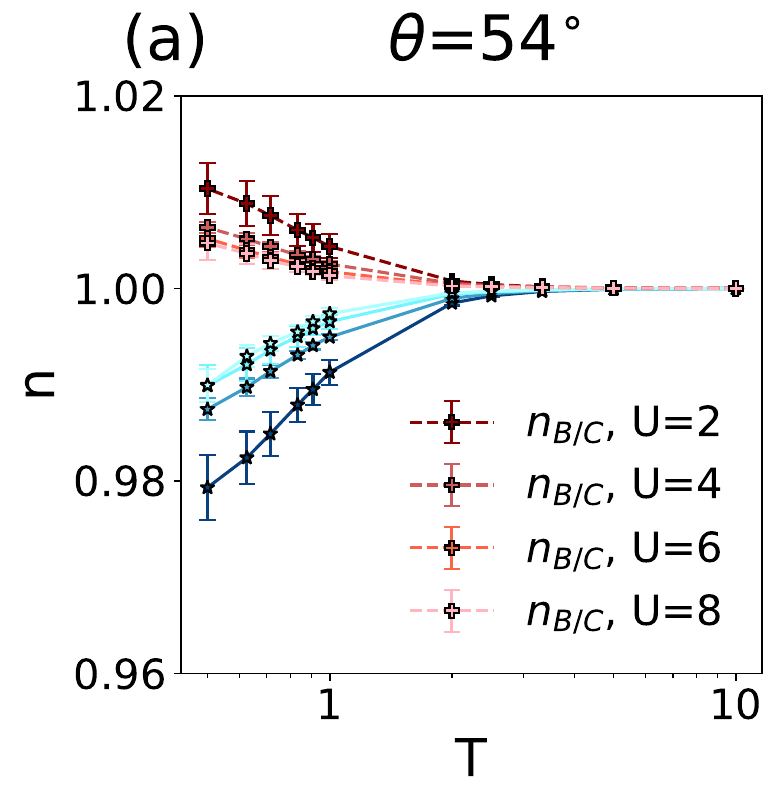}
\includegraphics[width=0.40\linewidth]{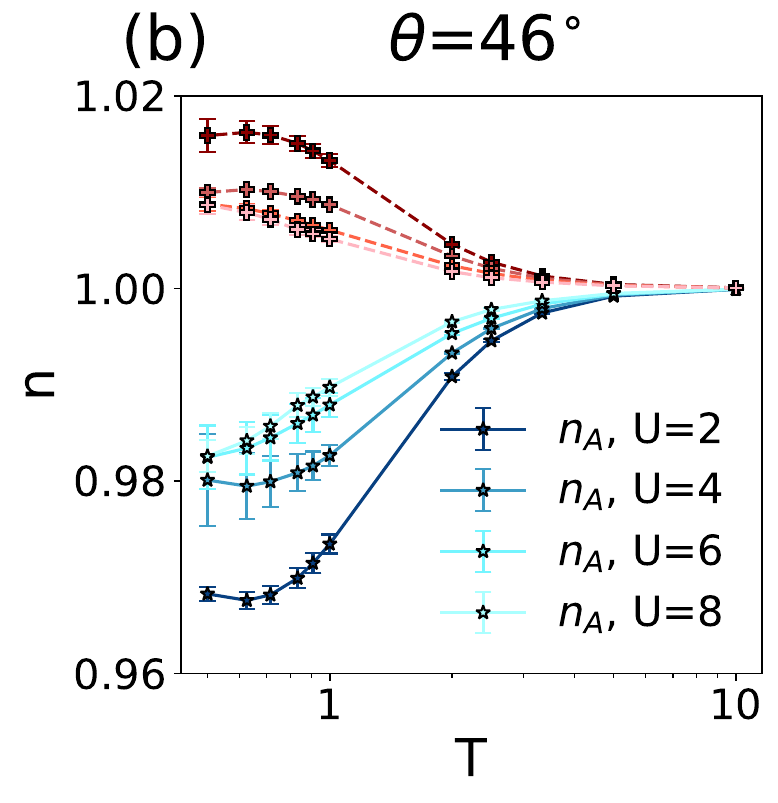}
\caption{Temperature dependence of the average particle number on sublattice $A$ and sublattices $B$/$C$ when the system is at half-filling. Error bars represent the standard error estimated from ten independent Monte Carlo simulations.}
\label{fig:fig3}
\end{figure}

When \(\theta = 60^\circ\), \(t_{AB}=t_{BC}=t_{AC}\), sublattices $A$, $B$, and $C$ are indistinguishable and exhibit identical properties, so that at half-filling each sublattice has an average particle number of 1. As \(\theta\) is reduced, sublattice $A$ responds differently from sublattices $B$ and $C$, leading to \(n_A\neq n_{B/C}\) as detailed in Fig.~\ref{fig:fig3}. At low temperatures and small \(U\), the number density imbalance between these two classes of sublattices becomes more pronounced, with $n_A$ exhibiting a larger deviation from \blue{unity}; in this kinetic-energy–dominated regime, variations in hopping amplitudes exert a stronger influence on the site occupations, thereby amplifying sublattice differentiation. As \(U\) increases, strong-correlation effects emerge on all sublattices, suppressing this differentiation. It is noteworthy that, despite the clear sublattice-dependent variations, the whole lattice half-filling condition is maintained, since 
\(\frac{1}{3}\bigl[n_A + 2\,n_{B/C}\bigr] \approx 1\).

\subsection{Conducting Properties}
\label{subsec:conductivity}

Double occupancy $D$ quantifies the likelihood of a lattice site being simultaneously occupied by spin-up and spin-down electrons, \(D=0.25\) at \(U=0\), indicating maximal double occupancy. This enables electrons to delocalize and the system exhibits metallic behavior. When \(D = 0\) as $U \rightarrow \infty$, the energy cost of adding an opposite-spin electron to an already occupied site becomes prohibitive, suppressing double occupancy and localizing electrons, thereby driving the system into an insulating state. ~\\ 

\begin{figure} [t!]
\includegraphics[width=0.323\linewidth]{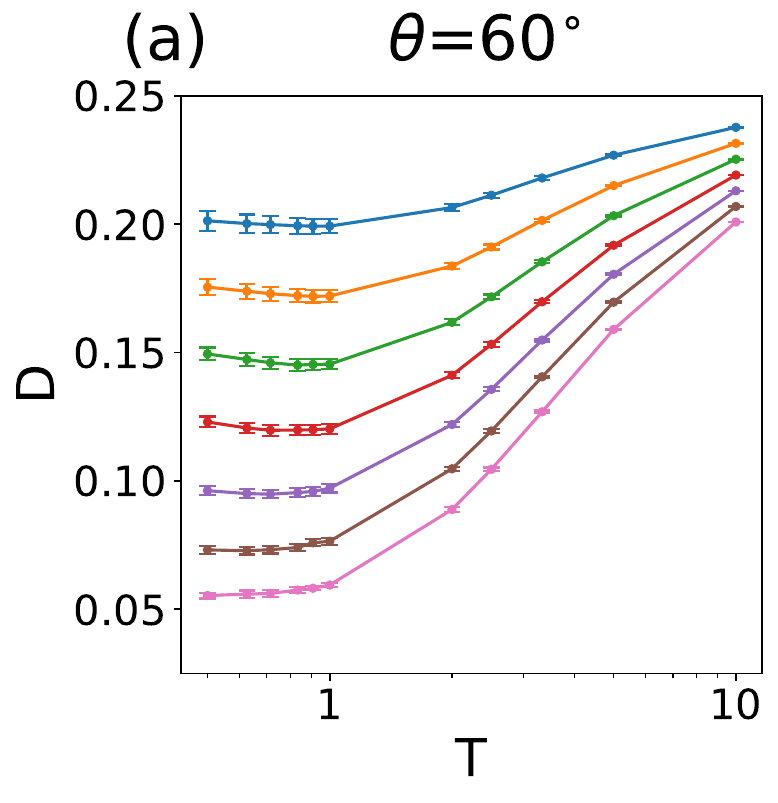}
\includegraphics[width=0.323\linewidth]{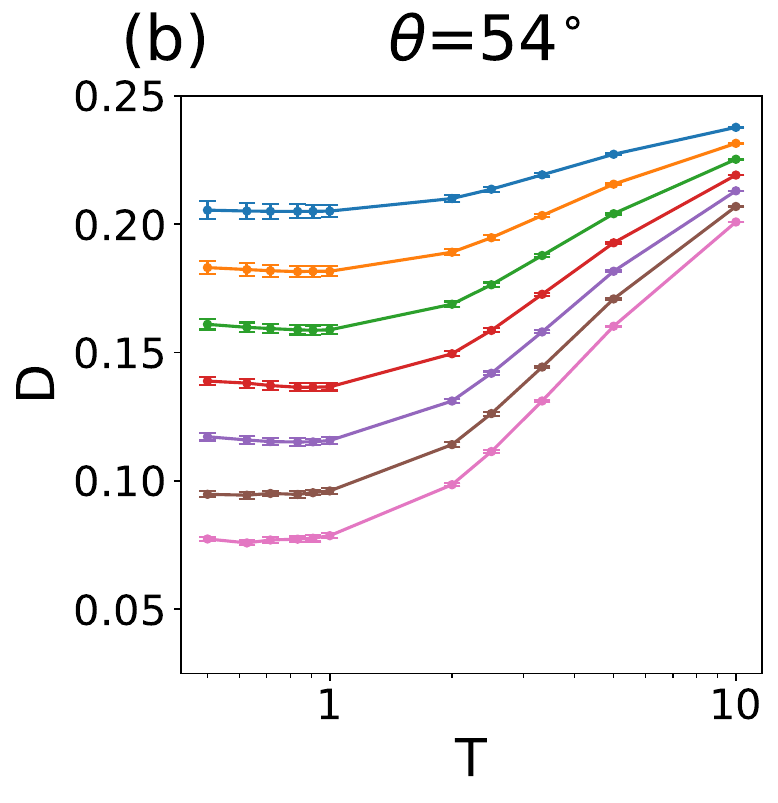}
\includegraphics[width=0.323\linewidth]{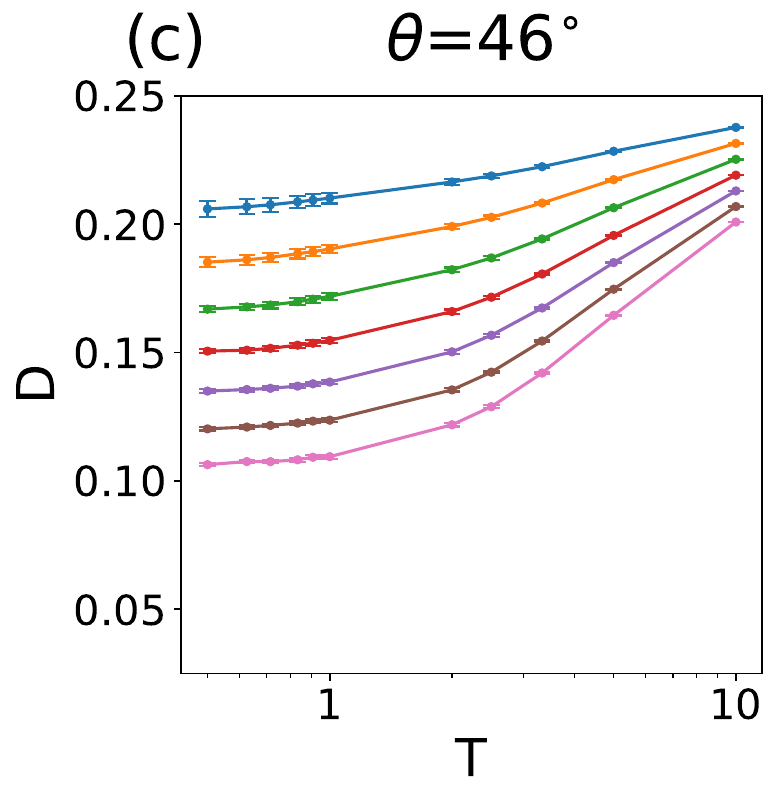}
\includegraphics[width=0.323\linewidth]{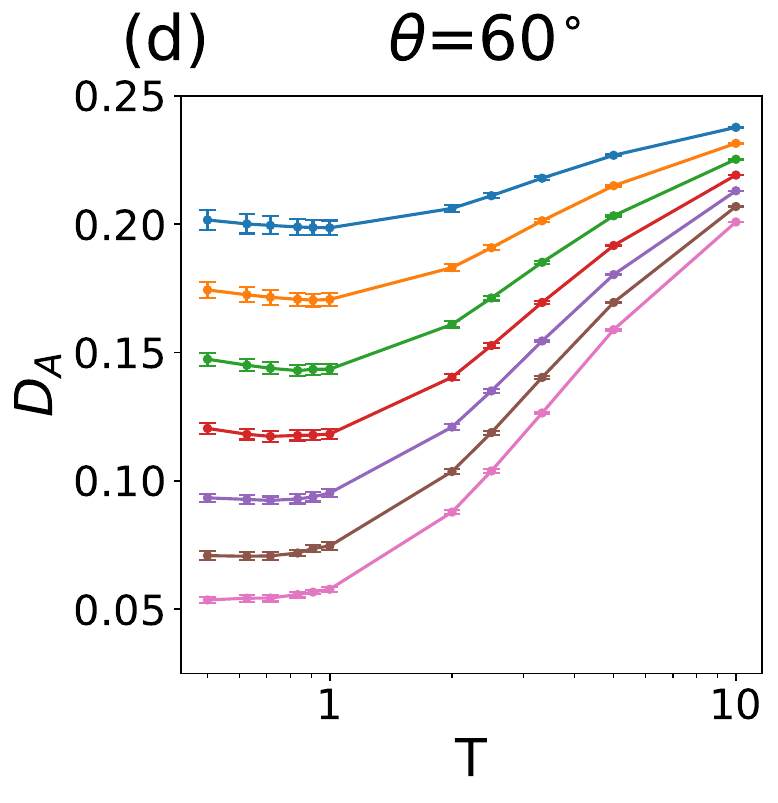}
\includegraphics[width=0.323\linewidth]{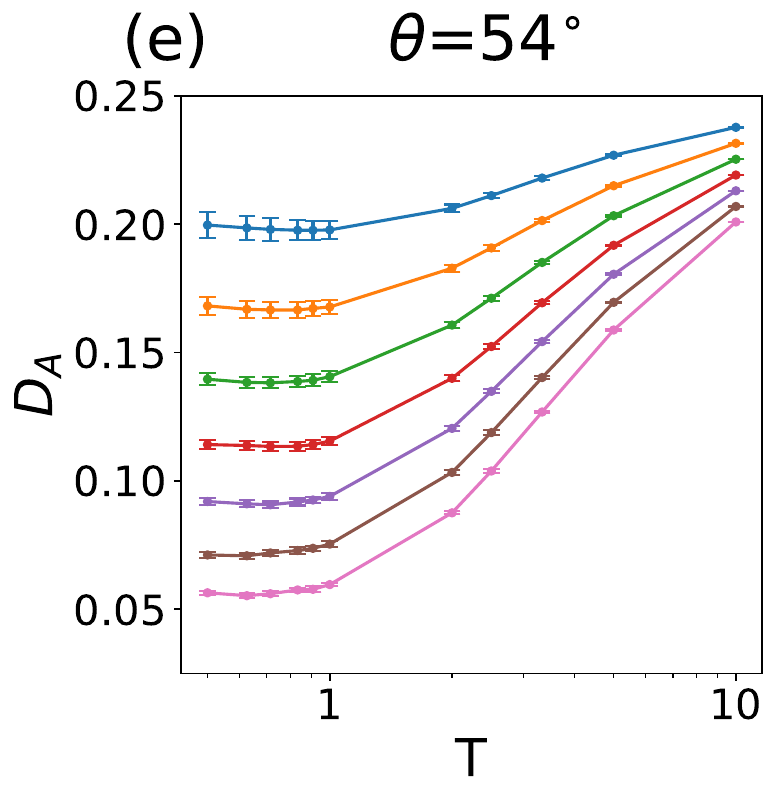}
\includegraphics[width=0.323\linewidth]{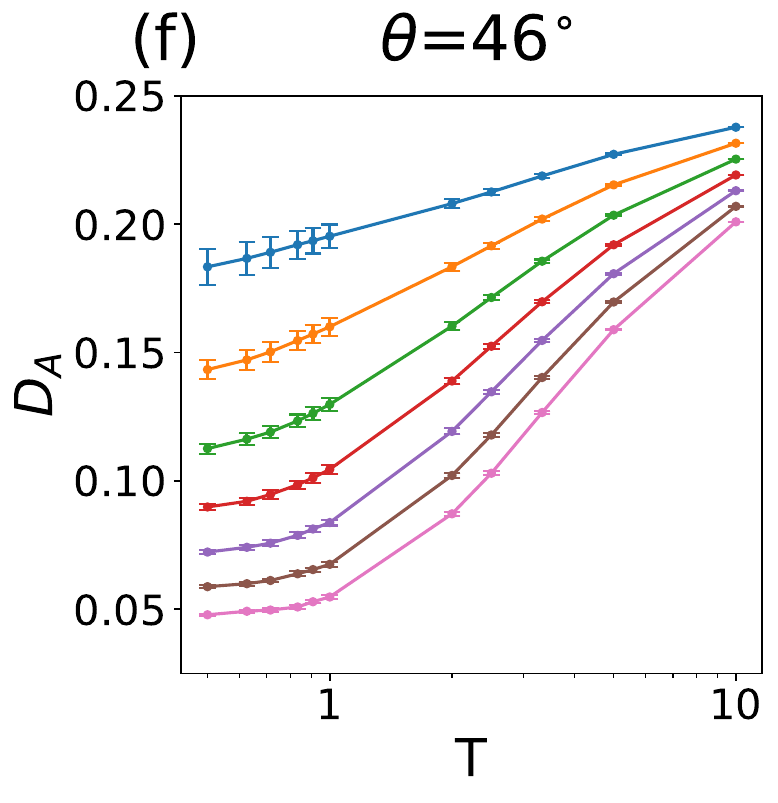}
\includegraphics[width=0.323\linewidth]{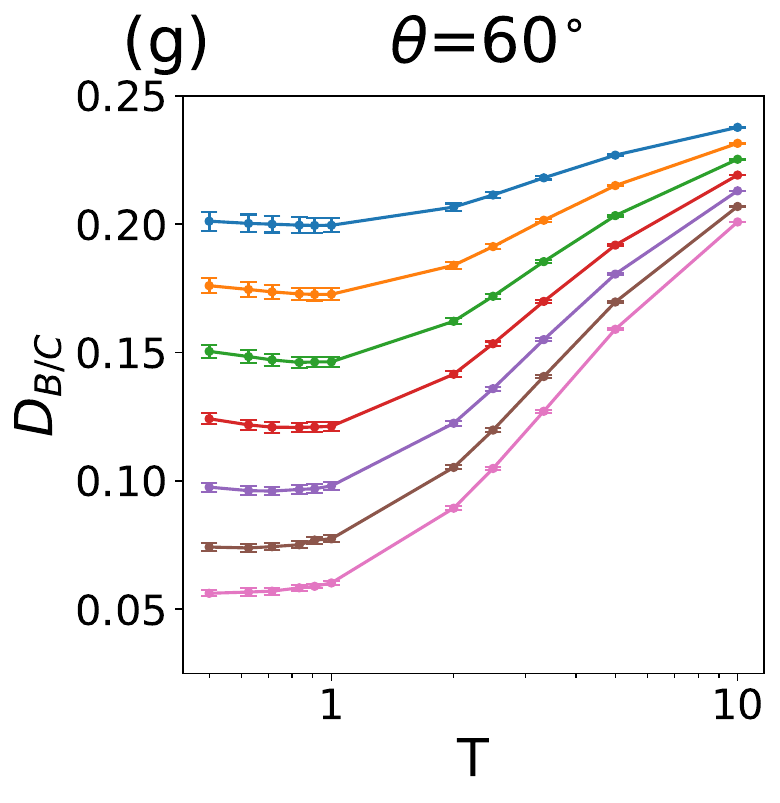}
\includegraphics[width=0.323\linewidth]{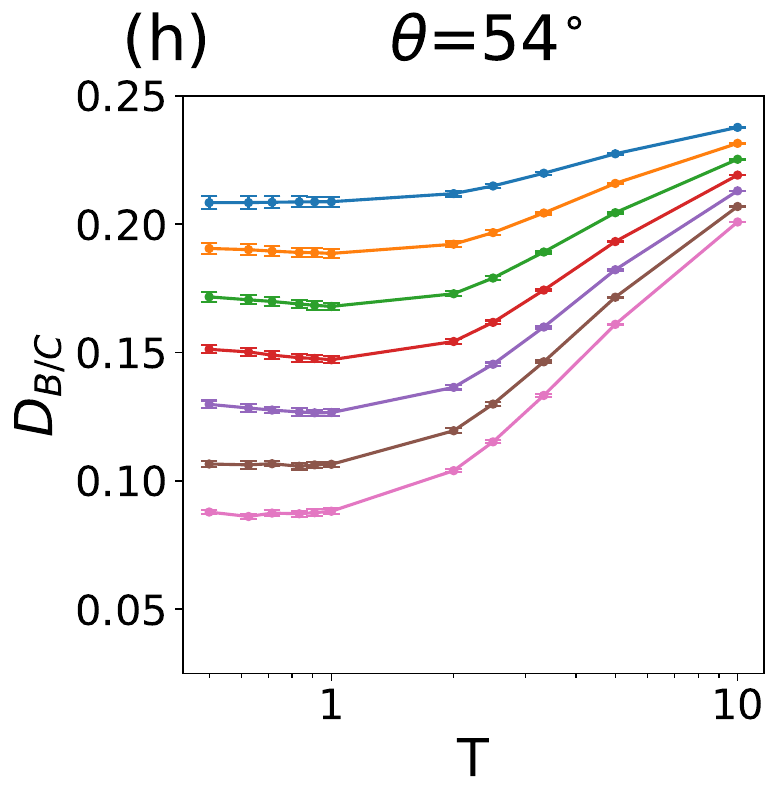}
\includegraphics[width=0.323\linewidth]{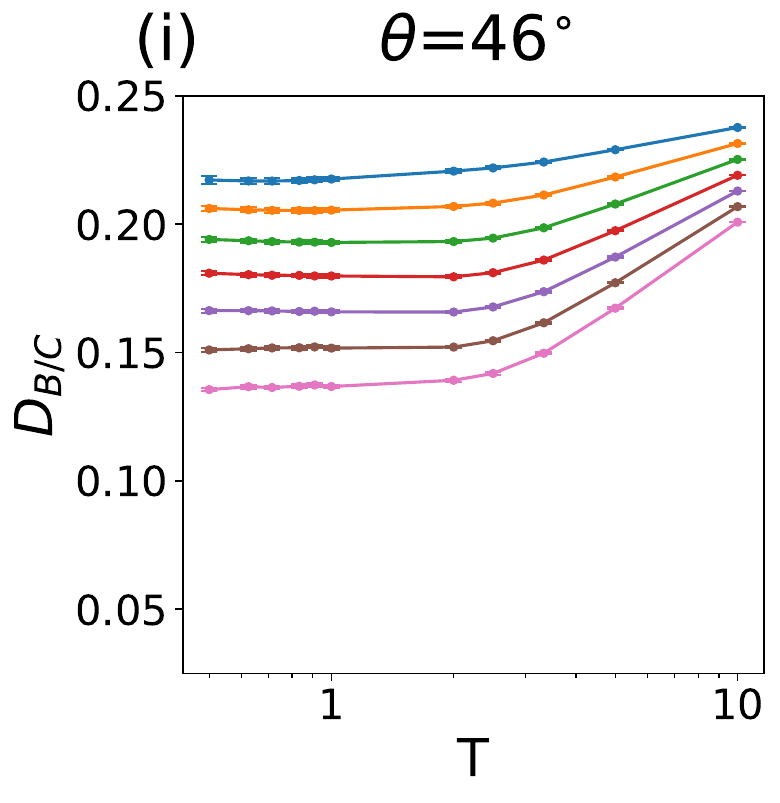}
\caption{Temperature dependence of the double occupancy \(D\) at half-filling for different interaction strengths \(U\). Top (a-c), middle (d-f), and bottom (g-i) panel show the average double occupancy of the whole lattice, sublattice $A$, and sublattice $B$/$C$, respectively. The color coding for the \(U\) values are the same as that of Fig. \ref{fig:fig2}. Error bars represent the standard error estimated from ten independent Monte Carlo simulations.}
\label{fig:fig4}
\end{figure}

Figure~\ref{fig:fig4}(a–c) shows the temperature dependence of the double occupancy \(D\) for the whole lattice. Remarkably, as \(\theta\) decreases, \(D\) at large \(U\) increases (e.g.\ for \(U=8\), \(D\) rises from \(\approx0.05\) at \(\theta=60^\circ\) to \(\approx0.11\) at \(\theta=46^\circ\)). 
As shown in Fig. ~\ref{fig:fig4}(d–f), \(D_{A}\) remains essentially unchanged when \(\theta\) is reduced from \(60^\circ\) to \(54^\circ\), but then shows obvious fall at \(\theta=46^\circ\) with an even stronger temperature dependence toward zero, indicating that sublattice $A$ tends toward Mott insulating state with suppressed double occupancy at \(\theta=46^\circ\). As shown in Fig. ~\ref{fig:fig4}(g–i), the $B$/$C$ sublattice resembles the whole lattice behavior: \(D_{B/C}\) increases as \(\theta\) decreases for a fix $U$. For \(\theta=46^\circ\), the weak temperature dependence of \(D\) in the low-temperature regime suggests that sublattice $B$/$C$ exhibits a stronger metallic behavior as compared to sublattice $A$. One may also consider the change of double occupancy with respect to temperature, in which $dD/dT < 0$ typically indicates metallic behavior, suggesting that as the temperature decrease, the double occupancy \(D\) tends to increase. On the other hand, \(dD/dT > 0\) usually corresponds to insulating behavior. Nevertheless, this is not always the case. Previous DQMC studies have shown that in kagome lattices, \(dD/dT\) does not accurately locate the transition from metallic to insulating states. \cite{medeiros2023thermodynamic}.~\\

\begin{figure} [t!]
\includegraphics[width=0.323\linewidth]{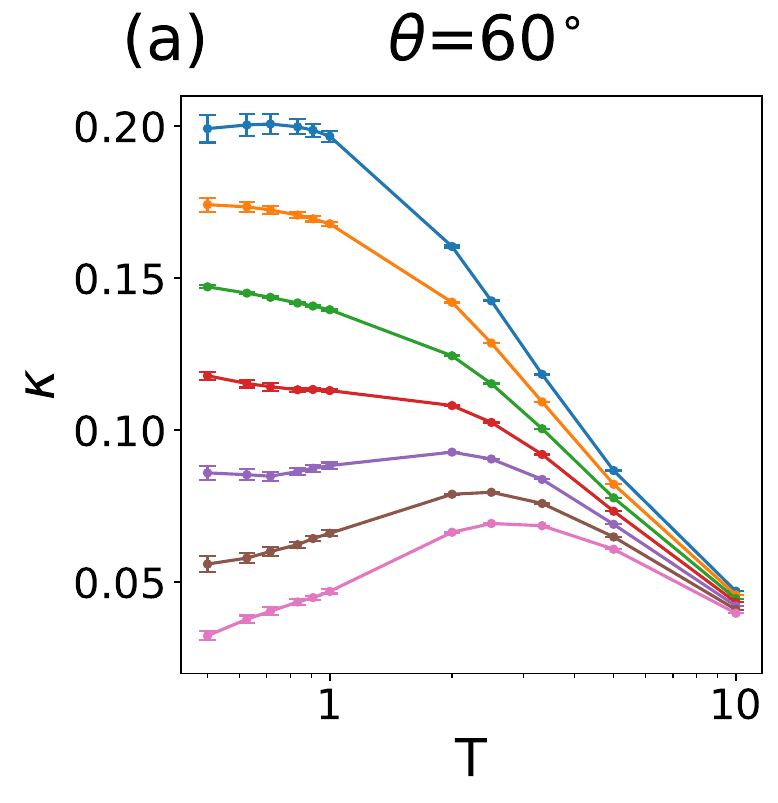}
\includegraphics[width=0.323\linewidth]{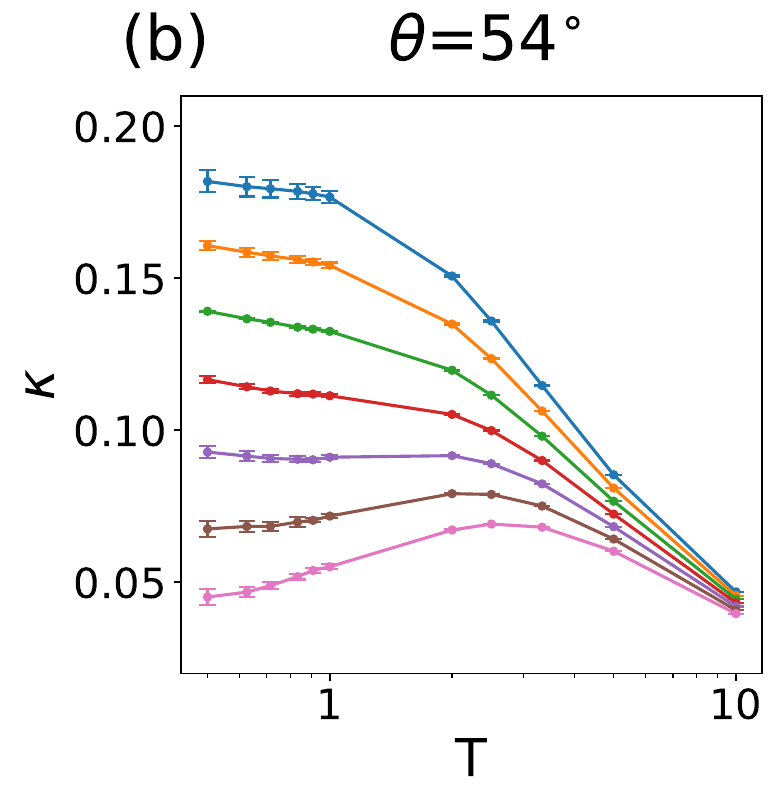}
\includegraphics[width=0.323\linewidth]{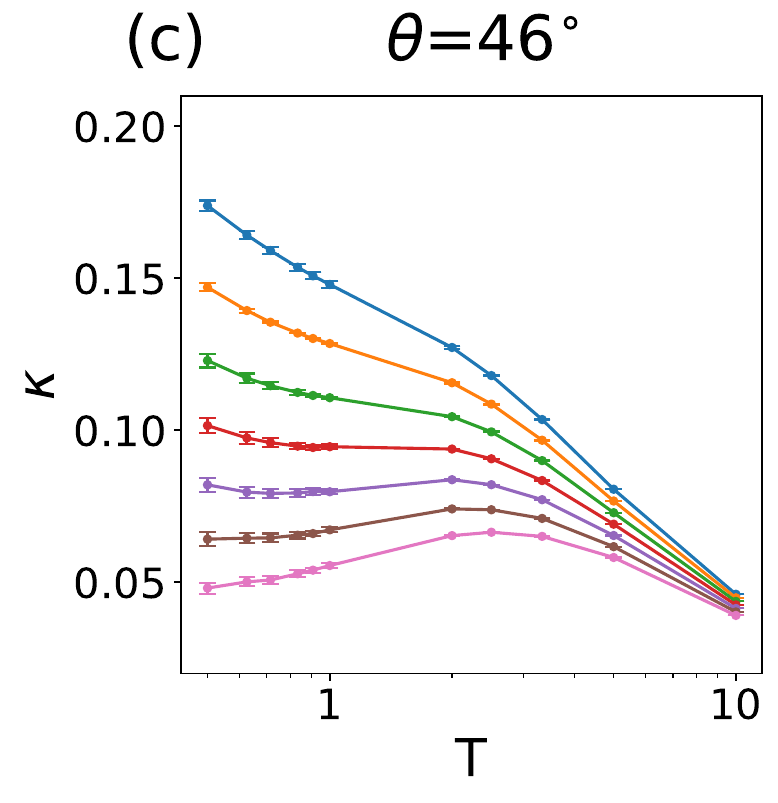}
\includegraphics[width=0.323\linewidth]{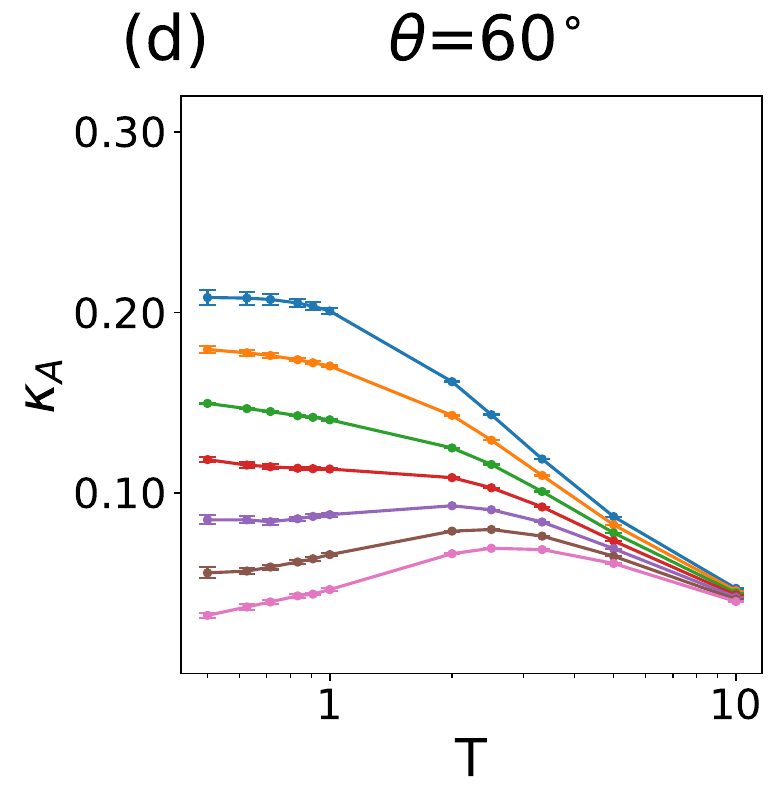}
\includegraphics[width=0.323\linewidth]{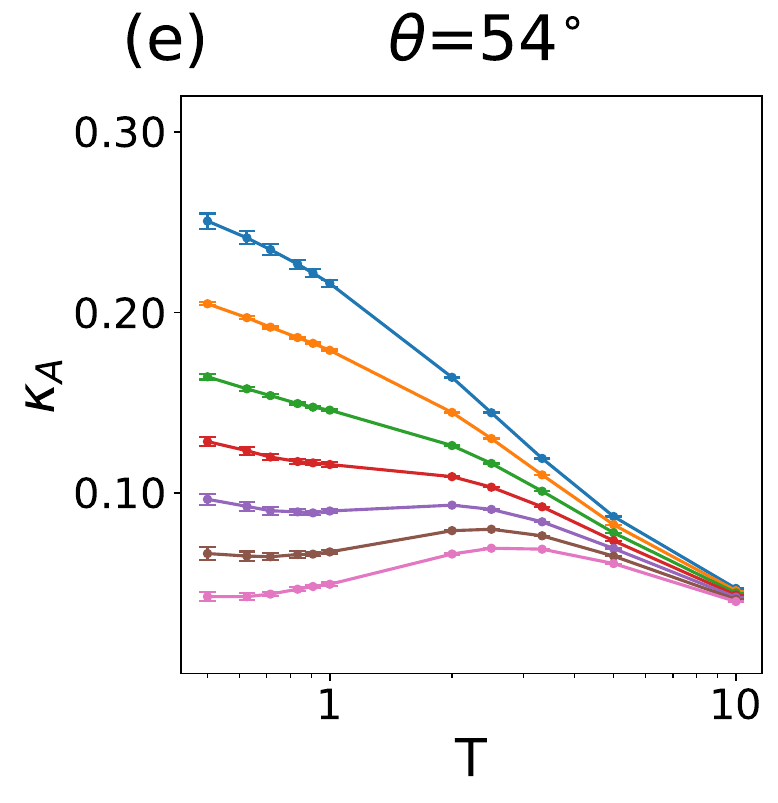}
\includegraphics[width=0.323\linewidth]{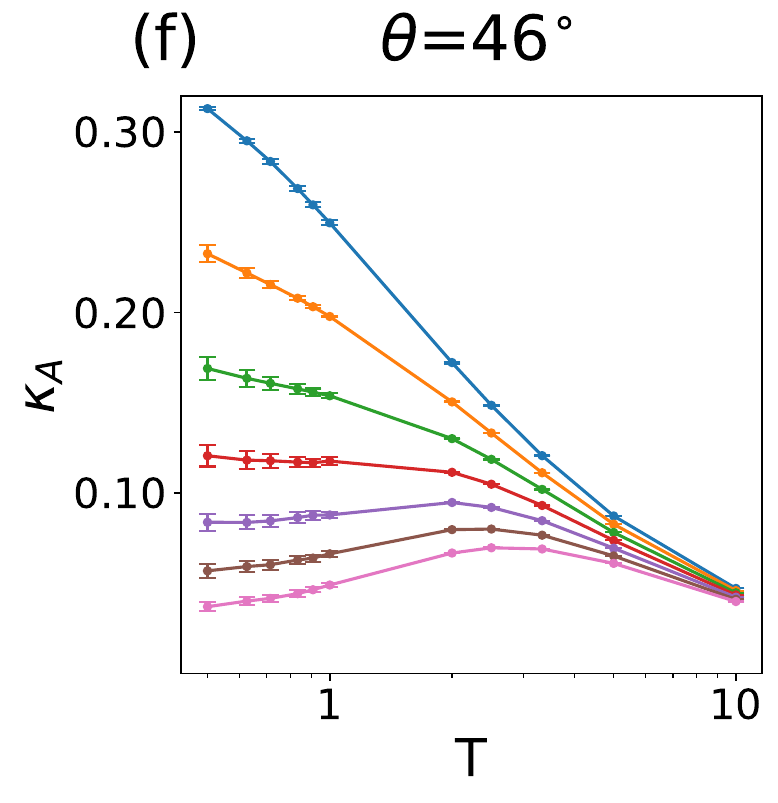}
\includegraphics[width=0.323\linewidth]{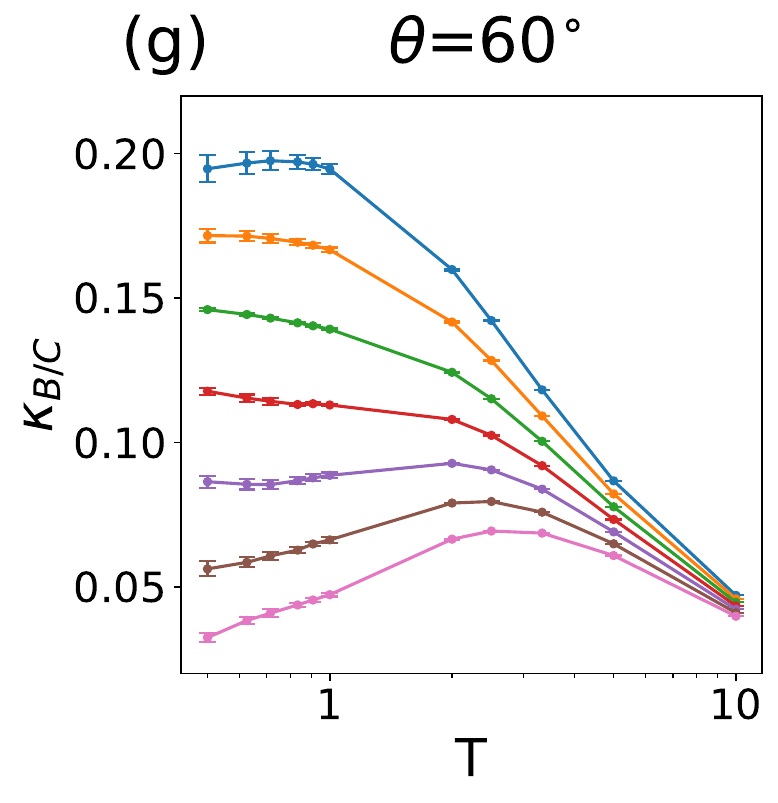}
\includegraphics[width=0.323\linewidth]{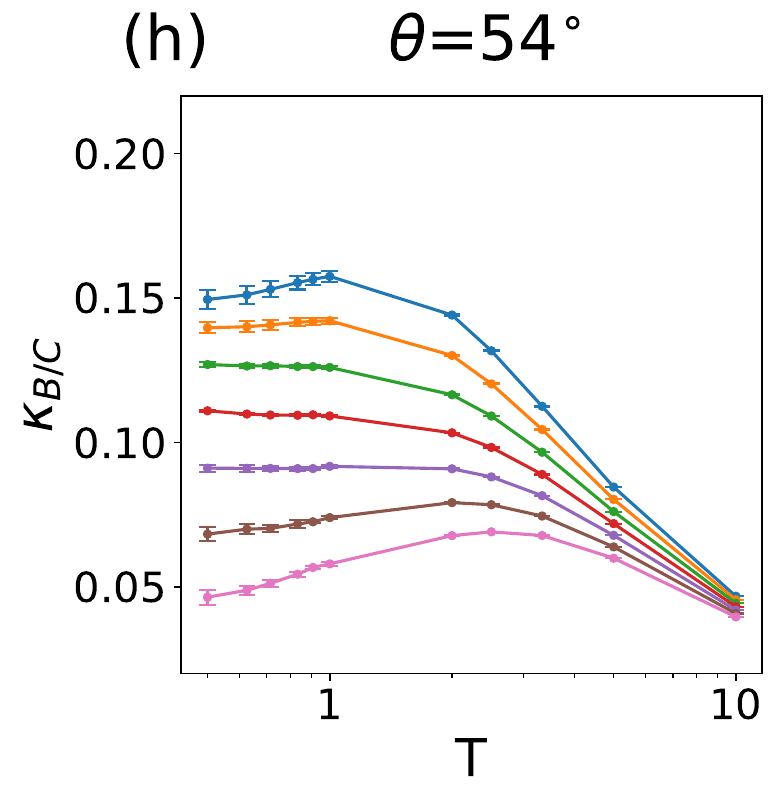}
\includegraphics[width=0.323\linewidth]{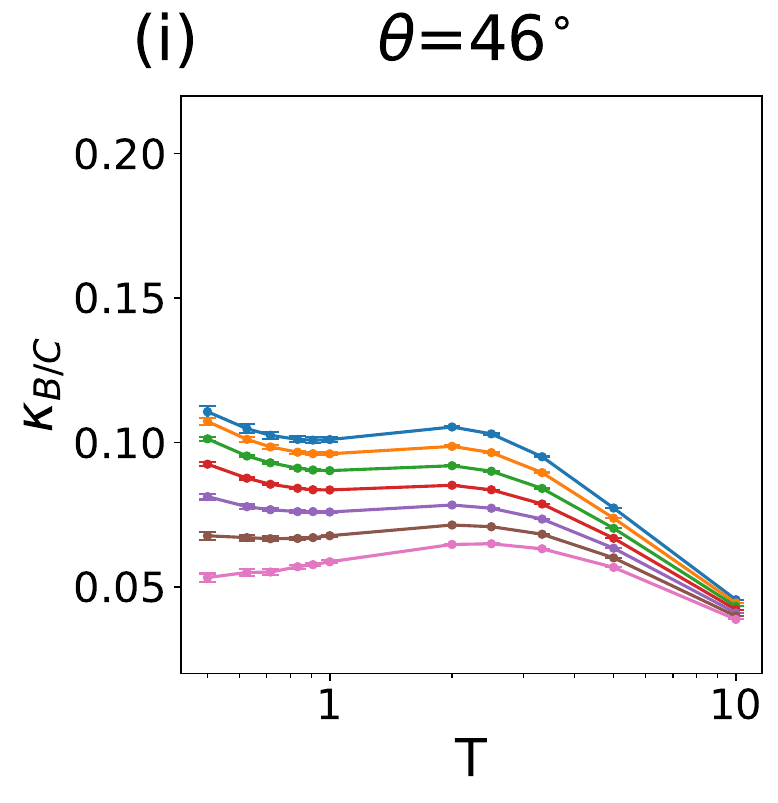}
\caption{Temperature dependence of the electronic compressibility $\kappa$ at half-filling for various values of $U$, with the color coding for the $U$ values consistent with that in Fig.~\ref{fig:fig2}. Top (a–c), middle (d-f), and the bottom (g-i) panel shows the average $\kappa$ for the whole lattice, sublattice $A$, and sublattice $B/C$, respectively. Error bars indicate the standard error from 10 independent Monte Carlo simulations.}
\label{fig:fig5}
\end{figure}

\blue{Electronic compressibility, \(\kappa\), provides an alternative probe of the system’s charge response and a useful indicator of the metal--Mott-insulator transition in the kagome lattice~\cite{medeiros2023thermodynamic}. We have extracted the temperature dependence of the electronic compressibility for the whole lattice, sublattice $A$, and sublattice $B$/$C$, as shown in Figs.~\ref{fig:fig5}(a--c), (d--f), and (g--i), respectively. We emphasize that the finite-temperature magnitude of \(\kappa\) itself is not used as a direct criterion for distinguishing a metal from a Mott insulator. In a finite-size finite-temperature DQMC calculation, \(\kappa\) is not expected to vanish \blue{completely}; the Mott insulating behavior is instead inferred from the low-temperature extrapolation of \(\kappa(T)\).} As \(\theta\) is reduced from \(60^\circ\) to \(46^\circ\), the temperature dependence of \(\kappa\) in the large-\(U\) regime weakens. The decrease in $\kappa$ at low $T$ is less pronounced, resulting in a larger electronic compressibility. \blue{This suggests that the low-temperature charge response is enhanced under compression.} Specifically, as shown in Figs.~\ref{fig:fig5}(a) and (c), for $U = 8$, both $\theta = 60^\circ$ and $\theta = 46^\circ$ reach a similar peak value of $\kappa \approx 0.065$ at $T \approx 3$. However, upon further cooling, $\kappa$ decreases more significantly for $\theta = 60^\circ$, reaching approximately $0.03$ at low temperatures, whereas for $\theta = 46^\circ$, $\kappa$ remains higher at around $0.05$. Comparing sublattices, one observes that at small \(\theta\), \(\kappa_A(T)\) remains strongly temperature-dependent, whereas \(\kappa_{B/C}(T)\) is essentially temperature-independent, reaffirming their \blue{markedly different charge responses}.~\\

\begin{figure} [t!]
\includegraphics[width=0.323\linewidth]{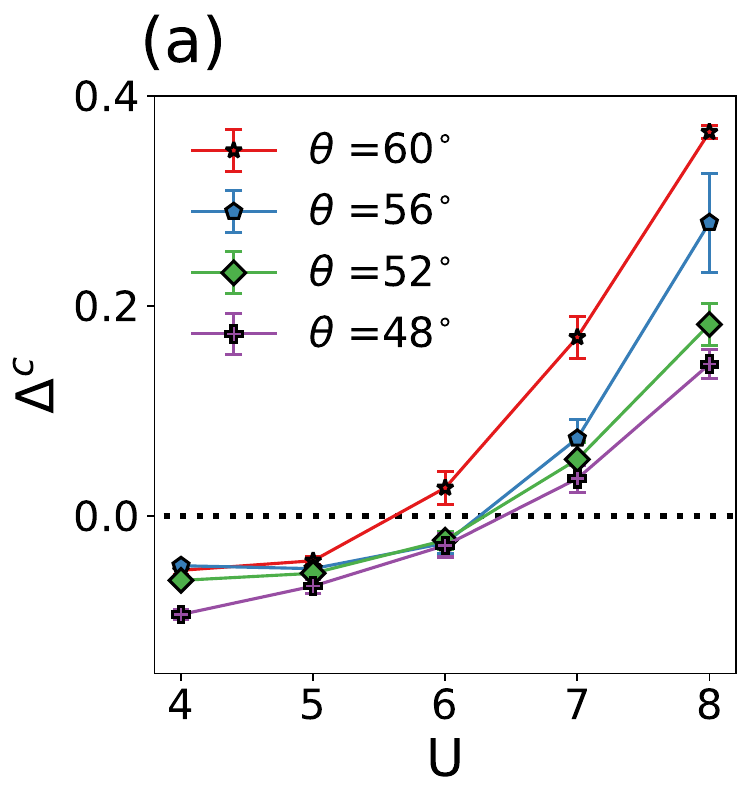}
\includegraphics[width=0.323\linewidth]{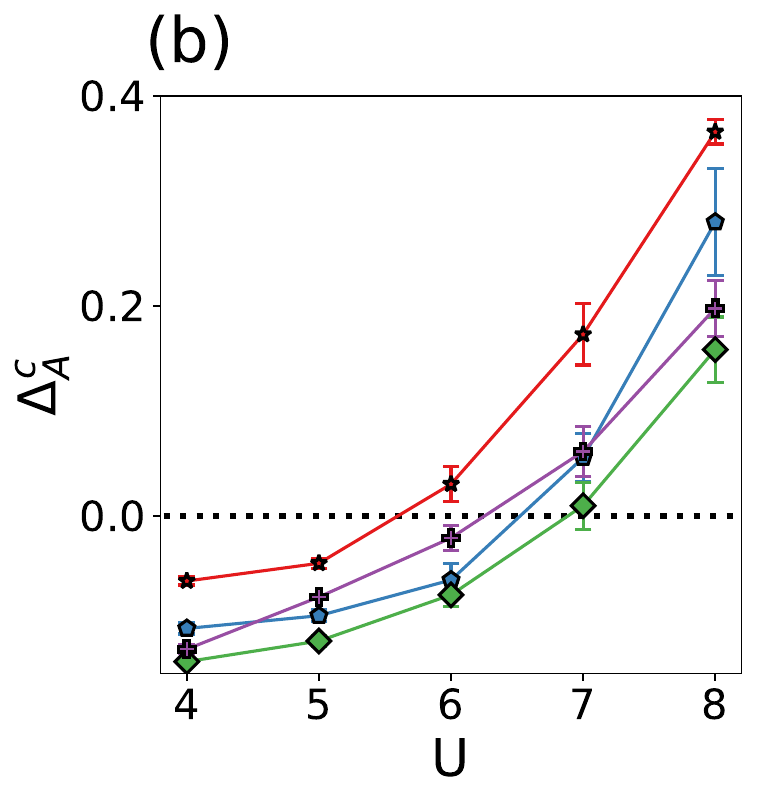}
\includegraphics[width=0.323\linewidth]{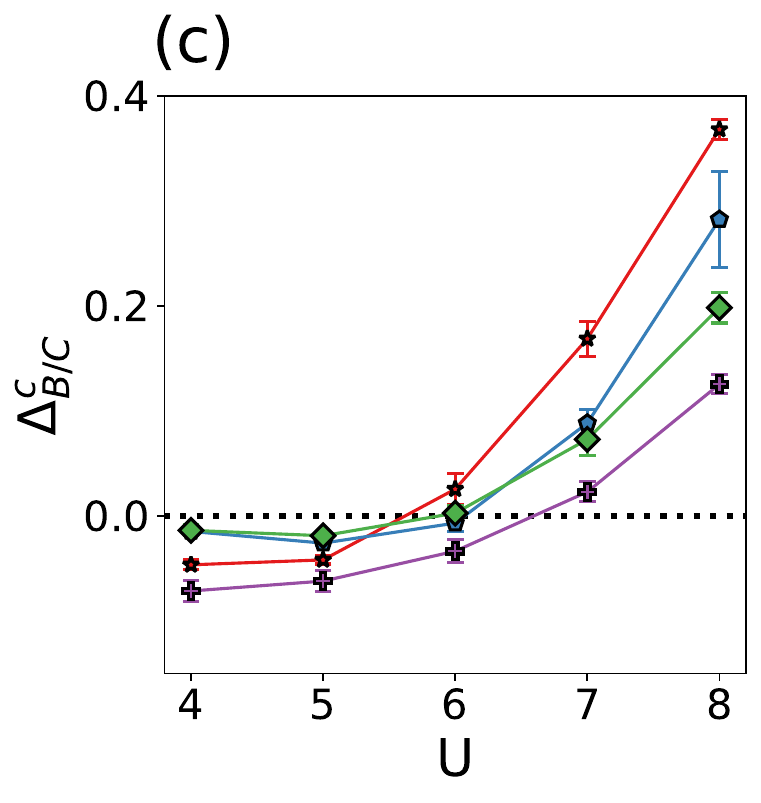}
\includegraphics[width=0.5\linewidth]{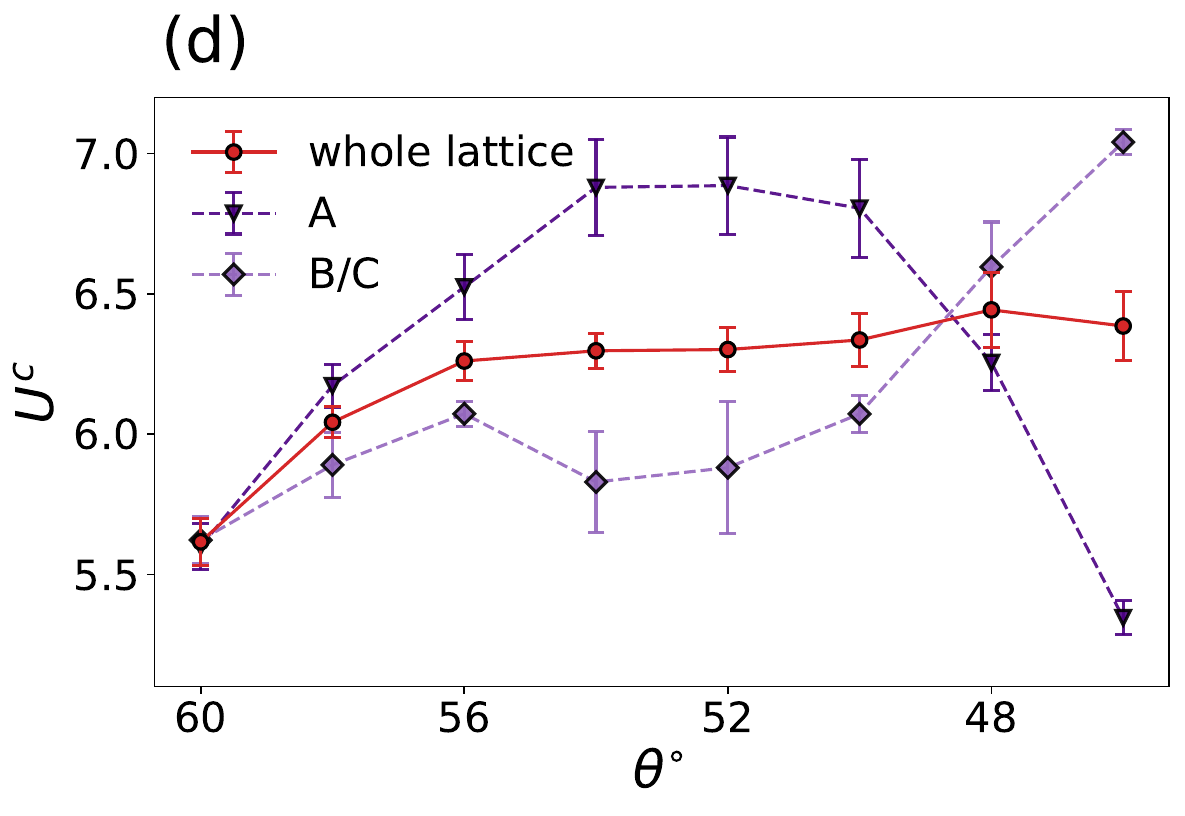}
\caption{\blue{(a–c) Effective charge activation scale \(\Delta^c\), extracted from the low-temperature fit \(\kappa(T)=k_1\exp(-\Delta^c/T)\), as a function of on-site interaction \(U\) for different values of \(\theta\) in the whole lattice, sublattice \(A\), and sublattices \(B\) or \(C\), respectively. The critical value \(U^c\) is estimated from \(\Delta^c=0\).} (d) $U^c$ as a function of $\theta$ at zero temperature, revealing a two-stage orbital-selective Mott transition. The error bars denote the standard errors derived from propagation of uncertainty in the fitting.}
\label{fig:fig6}
\end{figure}

\blue{Focusing on the low-temperature interval \(T<1\), we fit the compressibility to an activated form, $\kappa(T)=k_1\exp\!\left(-\Delta^c/T\right)$, where \(\Delta^c\) should be understood as an effective charge activation scale extracted from the finite-temperature compressibility, rather than as a single-particle band gap or a single-electron removal energy. \blue{The fitting procedure is detailed in Appendix \ref{sec:kappa_fit}}. In an incompressible Mott insulating regime, the low-temperature charge response is expected to be thermally activated, and a positive \(\Delta^c\) indicates that \(\kappa(T)\) is exponentially suppressed as \(T\to0\). In contrast, when the fitted \(\Delta^c\le 0\), the activated form does not describe a positive charge gap. We therefore interpret this regime as metallic or compressible, not as a state with a negative physical excitation gap. The metal--Mott boundary is estimated from the point where \(\Delta^c\) changes sign.} Figures~\ref{fig:fig6}(a–c) present the critical $U^c$ determined by $\Delta^c=0$ for the whole lattice, sublattice $A$, and sublattice $B/C$ at various values of $\theta$. To elucidate the sublattice differentiation, Fig.~\ref{fig:fig6}(d) shows the evolution of $U^c(\theta)$ for the whole lattice and different sublattices. The critical $U^c$ of the whole lattice exhibits a slight increase with decreasing $\theta$ and eventually saturates near $6.3\,t$. However, this seemingly smooth trend is actually the result of strong competition between $U^c_{A}$ and $U^c_{B/C}$. As $\theta$ decreases from $60^\circ$, $U^c_{A}$ rises rapidly while $U^c_{B/C}$ increases only gradually, indicating that when sublattice $A$ remains metallic, sublattices $B/C$ have already entered the Mott insulating regime, characteristic of an orbital-selective Mott phase associated with sublattices $B/C$~\cite{anisimov2002orbital,liebsch2004single,demedici2005orbital,koga2004orbital,yu2013orbital,herbrych2018spin,kim2022orbital,kim2024orbital}. As $\theta$ is further reduced to $\sim52^\circ$, $U^c_{A}$ drops sharply while $U^c_{B/C}$ begins to increase rapidly; at $\theta\approx49^\circ$, $U^c_{A}$ and $U^c_{B/C}$ intersect, and for $\theta$ below this value, $U^c_{B/C}$ remains higher than $U^c_{A}$ by about 30\% at $\theta=46^\circ$, revealing an orbital-selective Mott phase dominated by sublattice $A$. In summary, the system exhibits a two-stage orbital-selective Mott transition.~\\

It should be noted that when we refer to $U^{c}$ as marking the metal-Mott insulator transition of the whole lattice, we are describing a global measurement extracted from the lattice-averaged compressibility $\kappa$, while $U^{c}_{A}$ and $U^{c}_{B/C}$ are defined in a similar way by considering the sublattices. In experiment, since bulk probes such as DC transport effectively average over the unit cells, one may observe that the conductivity drops sharply near $U \approx U^{c}_{A}$ (or $U \approx U^{c}_{B/C}$) when one sublattice becomes nearly incompressible. However, a fully insulating behavior generally requires localization of electrons on all relevant sublattices. This is reminiscent of sublattice/site-selective correlation effects discussed in rare-earth nickelates and related systems~\cite{park2012site,layek2024correlated}. It is different from the case $U \gtrsim \max\{U^{c}_{A},\,U^{c}_{B/C}\}$, where both $\kappa_{A}$ and $\kappa_{B/C}$ are strongly suppressed at low $T$, so that the system becomes globally incompressible and the DC conductivity is expected to vanish in the $T\to 0$ limit. In cold-atom realizations of the Fermi–Hubbard model, local compressibility can be extracted from in-situ density profiles, providing a direct route to distinguish global responses from local crossovers~\cite{duarte2015compressibility}.
~\\

\subsection{Magnetic Properties}
\label{subsec:magenization}

The magnetic properties of the system are also a central focus of this study, as magnetism is often closely linked to conductivity. We divide the system  at $\beta=2$ into the $AB$ and $BC$ paths, the transverse spin correlation function $G^{x}(r)$ along each path is computed. Although the transverse and longitudinal spin correlation functions are theoretically equivalent, the statistical error of $G^{x}$ is significantly smaller than that of $G^{ z}$, making it more suitable for detailed analysis (see Appendix~\ref{sec:sz_sx}).~\\

Figures~\ref{fig:fig7}(a–c) show the dependence of $G^{x}$ on distance $r$ along the $AB$ path, which decays rapidly to zero, indicating the absence of long-range magnetic correlations and thus a paramagnetic state along the $AB$ direction. At $\theta=46^\circ$, $G^x(r=0)$ increases for all values of $U$. Since $S^x$ and $S^z$ exhibit similar behavior, it follows that $G^z(r=0)$ also increases across all $U$. Notably, $S^z = n_{\uparrow} - n_{\downarrow}$, so a larger $|S^z|$ indicates a higher tendencies of single occupancy at the corresponding site
(i.e., $n_{\uparrow} \to 1$, $n_{\downarrow} \to 0$, or vice versa), consistent with our previous observation that sublattice $A$ at $\theta=46^\circ$ suppresses double occupancy and exhibits reduced metallicity. In addition, at $\theta=60^\circ$, $G^x(r=1)$ shows a negative nearest-neighbor correlation, likely reflecting extremely short-range magnetic correlations arising from geometric frustration among the three sublattices in the kagome lattice. A similar feature is observed in Fig.~5(a) of Ref.~\cite{lima2023magnetism} for $t'/t=1$. As $\theta$ decreases, the frustration between sublattices is reduced, and the nearest-neighbor correlation approaches zero.~\\

Figures~\ref{fig:fig7}(d–f) present $G^x$ along the $BC$ path. As $\theta$ decreases, $G^x(r=0)$ decreases for all values of $U$, indicating that sublattices $B$ and $C$ tend to favor double occupancy and exhibit enhanced metallic. For $\theta \geq 54^\circ$, $G^x(r)$ decays rapidly over the entire $r$ range, reflecting short-range antiferromagnetic correlations. In contrast, at $\theta = 46^\circ$, $G^x$ develops long-range correlations with alternating sign. Furthermore, as shown in Fig.~\ref{fig:fig7}(g), the local moments of sublattices $B$ and $C$ are comparable in magnitude, \blue{supporting the development of antiferromagnetic correlations along the $BC$ path}.~\\

\begin{figure} [t!]
\includegraphics[width=0.323\linewidth]{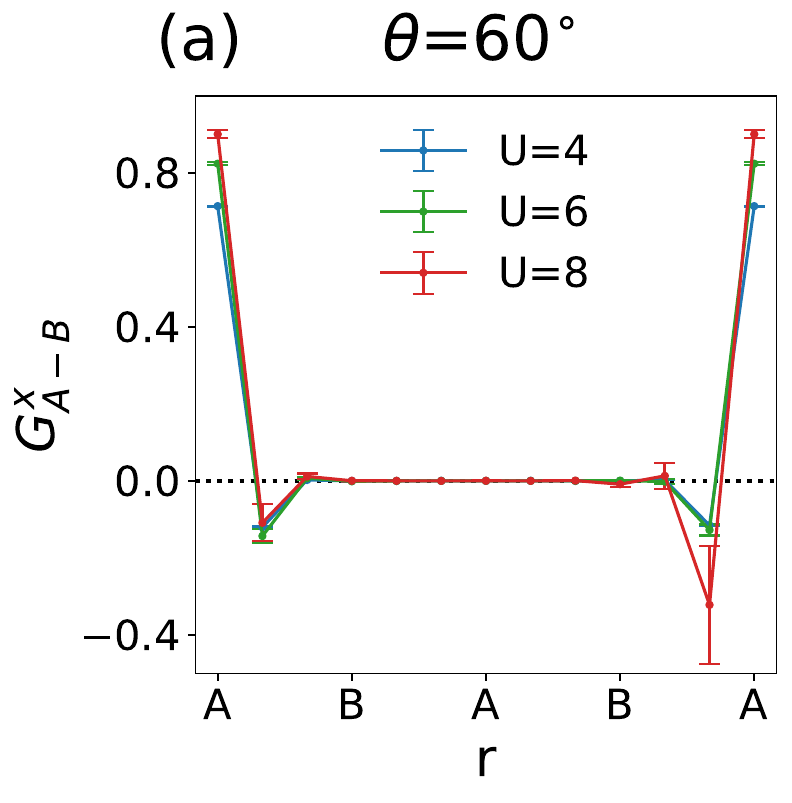}
\includegraphics[width=0.323\linewidth]{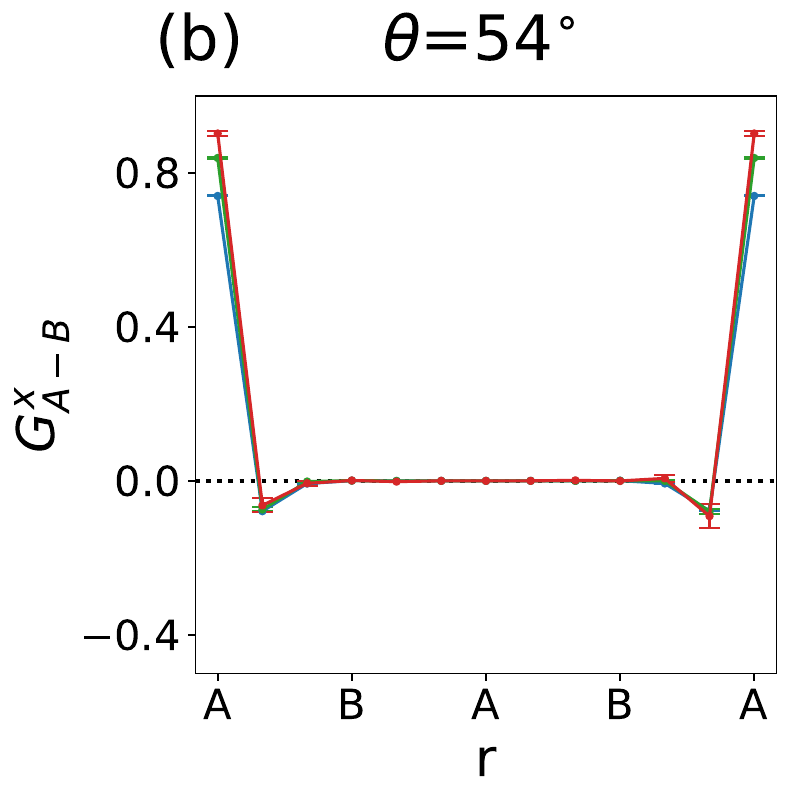}
\includegraphics[width=0.323\linewidth]{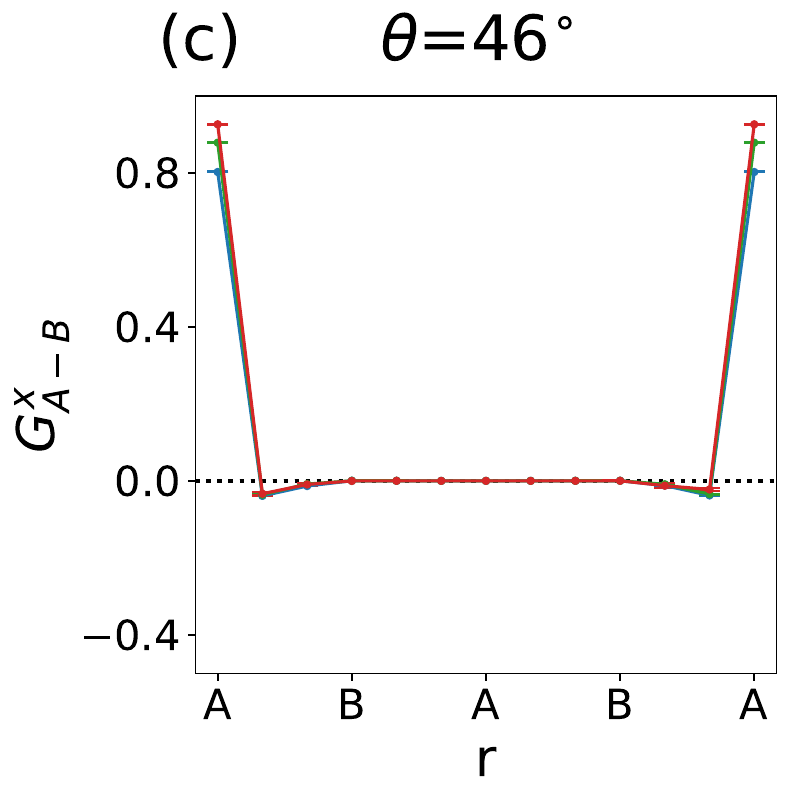}
\includegraphics[width=0.323\linewidth]{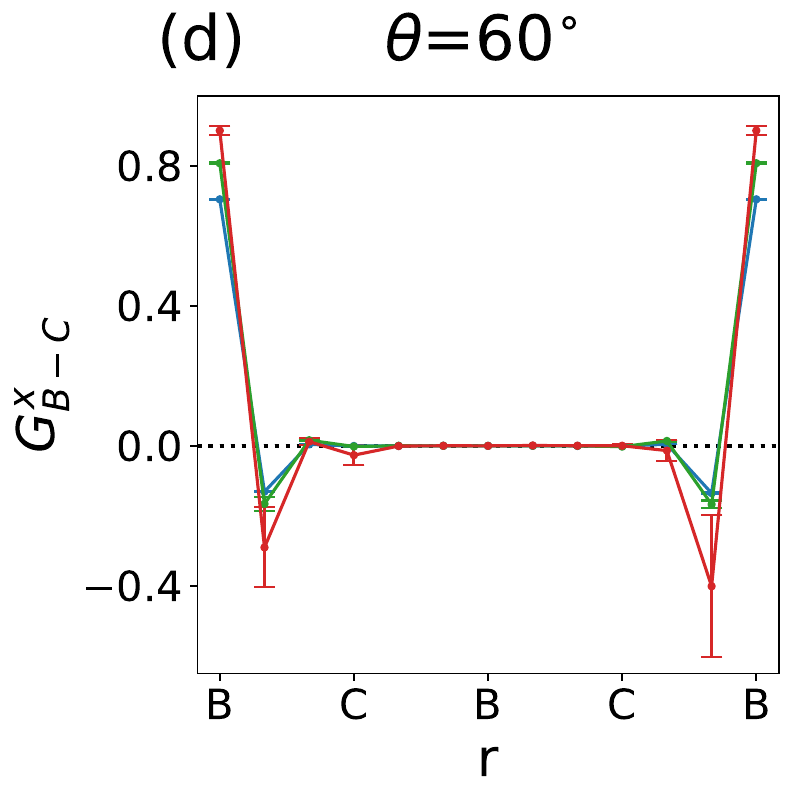}
\includegraphics[width=0.323\linewidth]{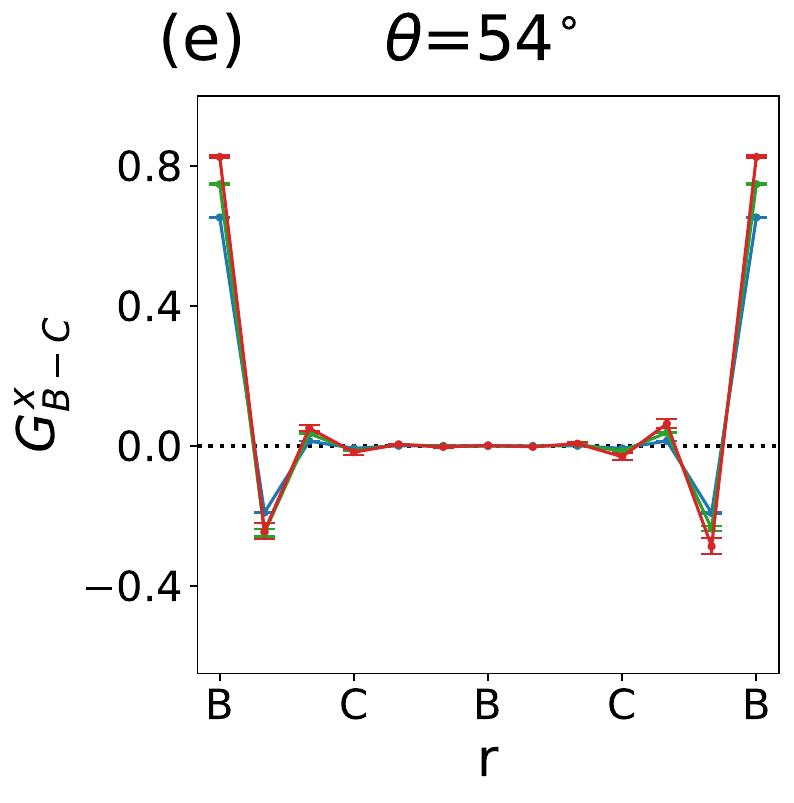}
\includegraphics[width=0.323\linewidth]{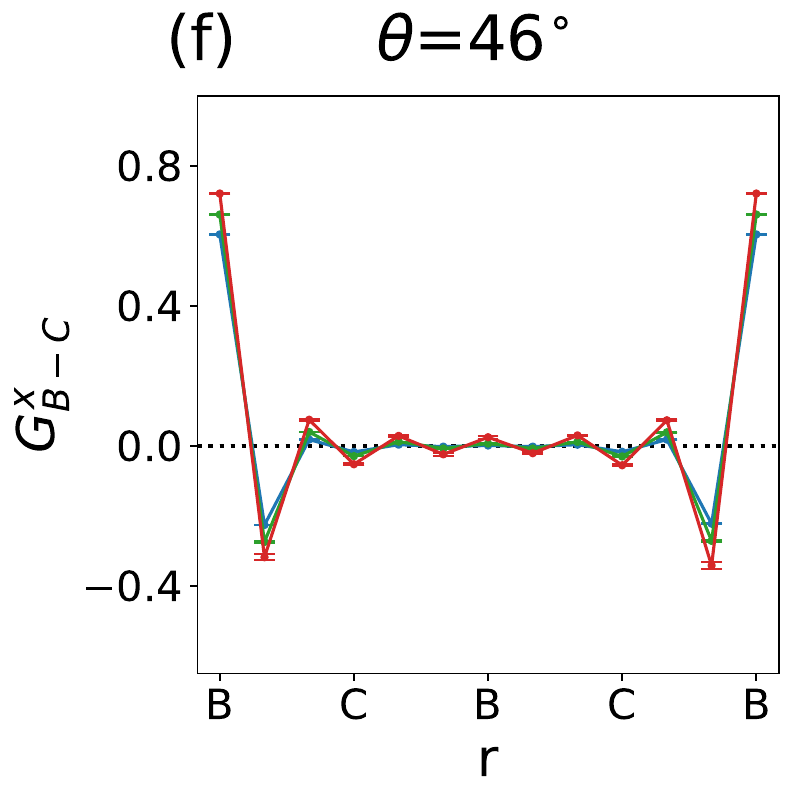}
\includegraphics[width=0.4\linewidth]{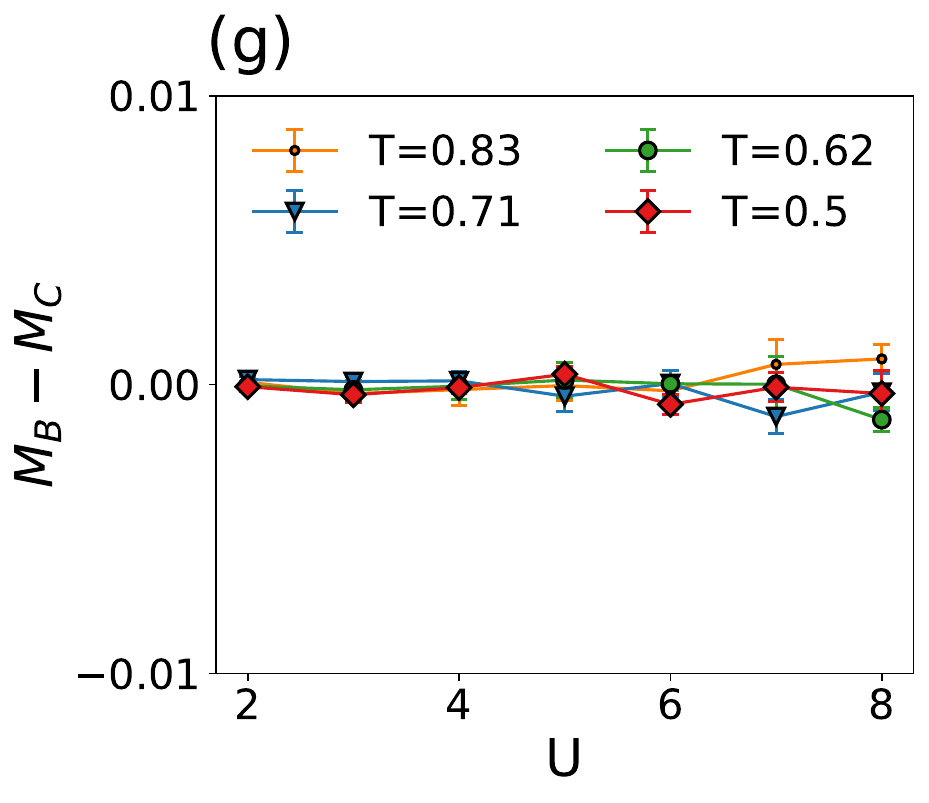}
\caption{(a–c) Transverse spin correlation function $G^{x}(r)$ along the $AB$ path at various $U$ for $\beta$=2. (d–f) The same function along the $BC$ path. (g) Difference between the local moments of sublattice $B$ and sublattice $C$ as a function of $U$ at $\theta=46^\circ$. Error bars represent the standard error from 10 independent Monte Carlo Markov chain measurements. }
\label{fig:fig7}
\end{figure}

\begin{figure} [t!]
\includegraphics[width=0.32\linewidth]{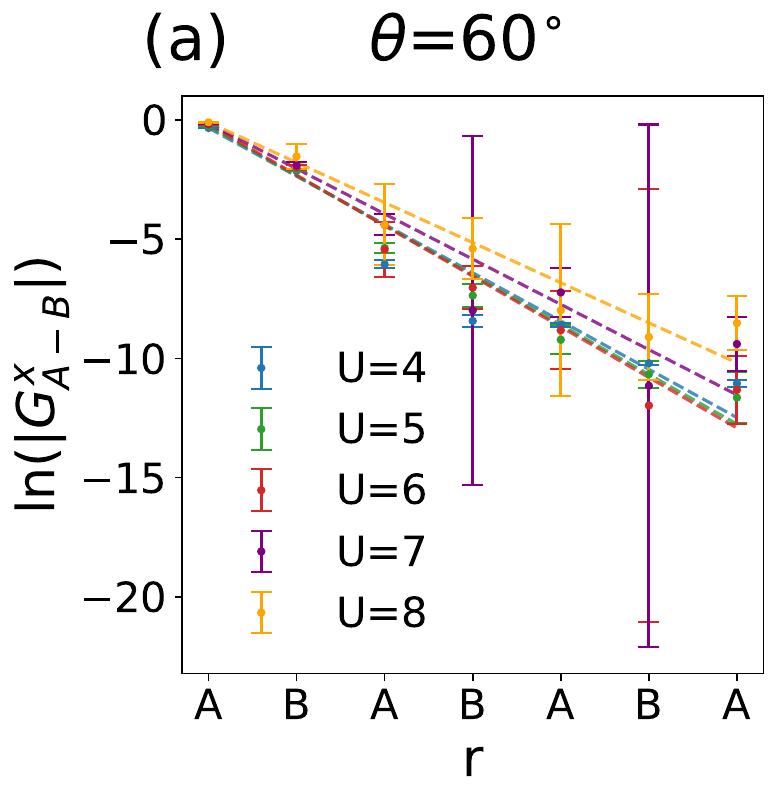}
\includegraphics[width=0.32\linewidth]{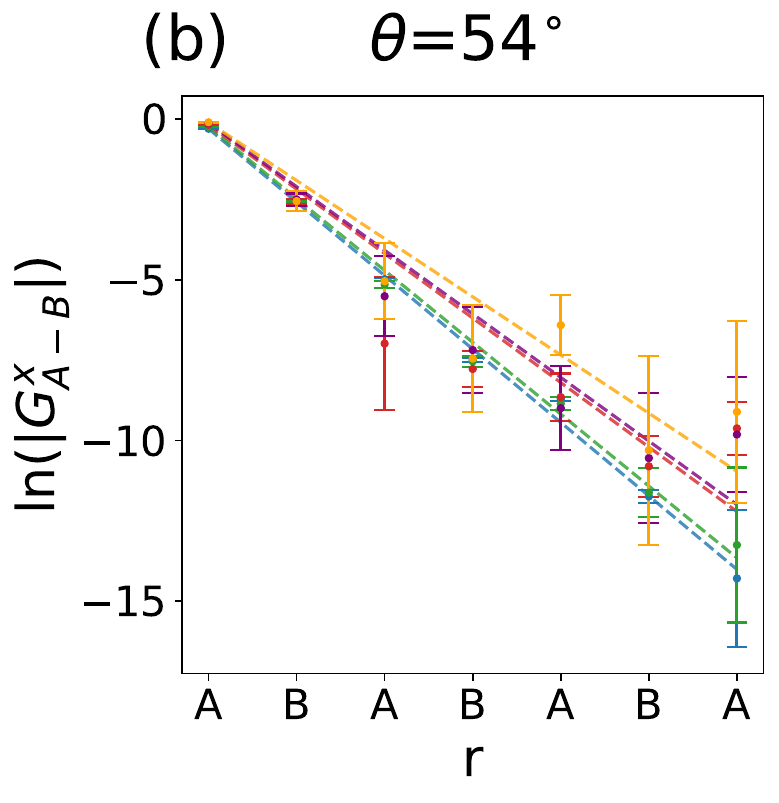}
\includegraphics[width=0.32\linewidth]{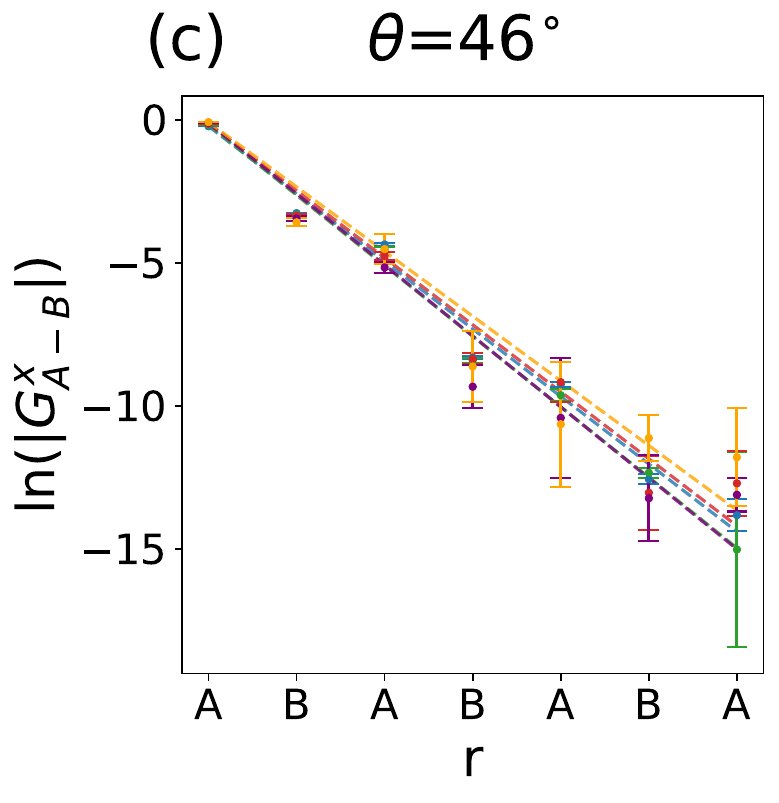}
\includegraphics[width=0.32\linewidth]{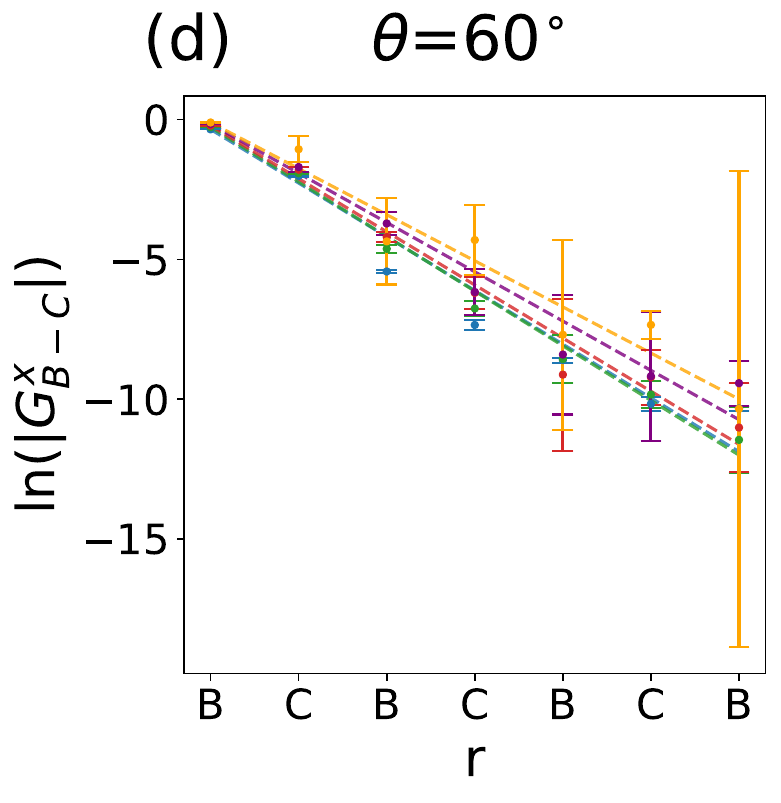}
\includegraphics[width=0.32\linewidth]{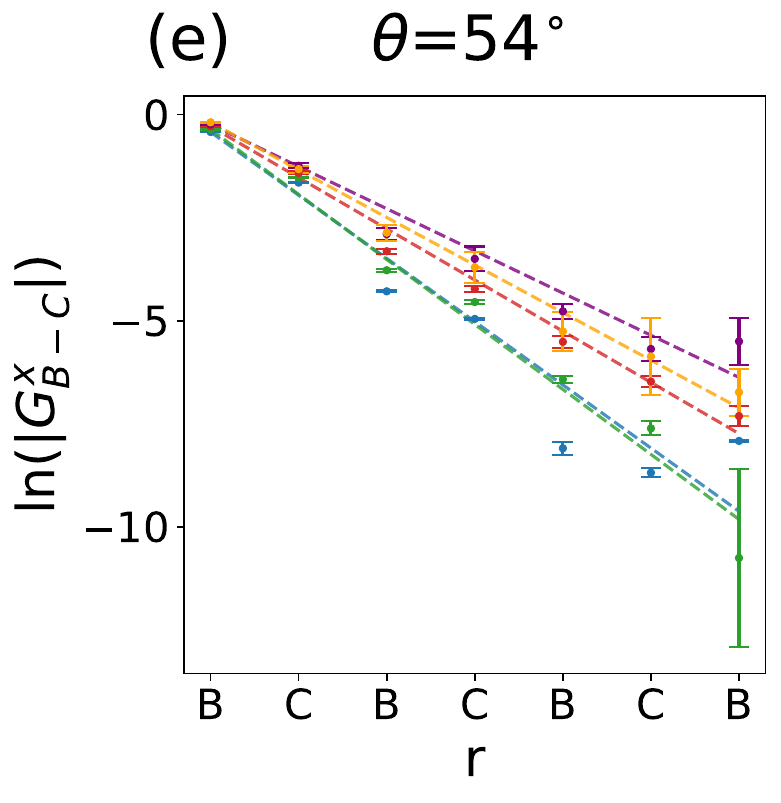}
\includegraphics[width=0.31\linewidth]{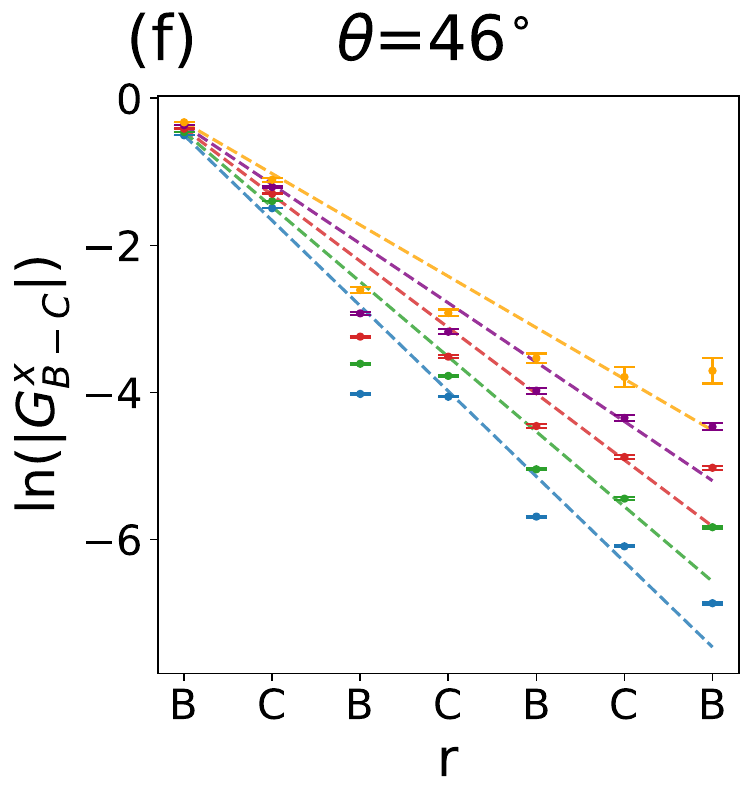}
\includegraphics[width=0.4\linewidth]{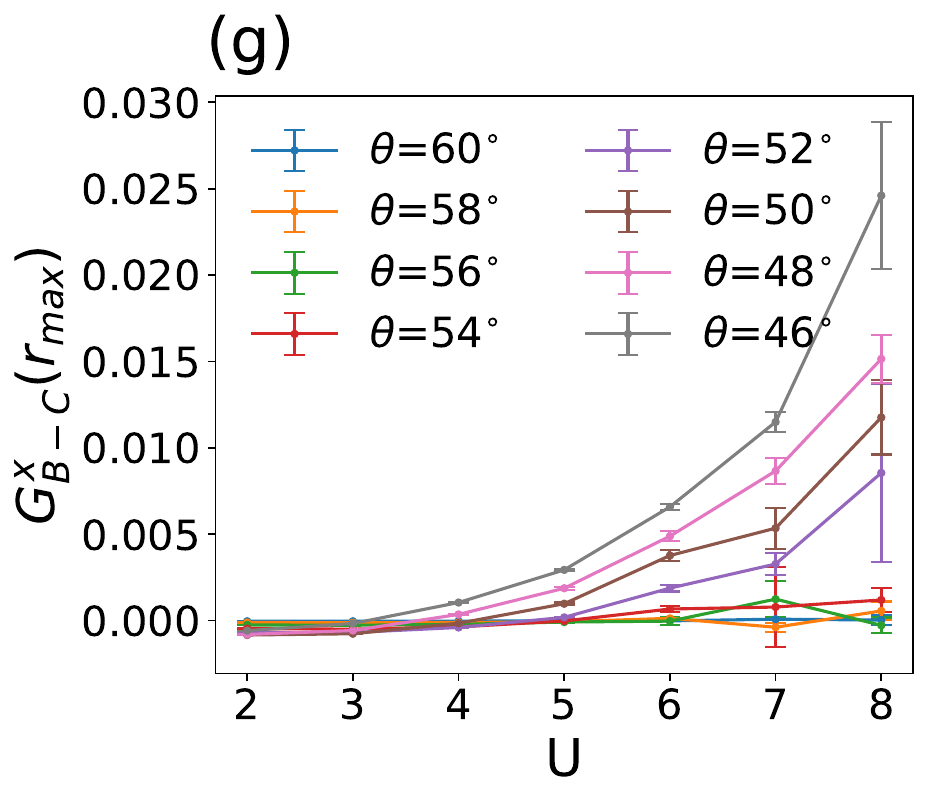}
\caption{(a--c) Logarithm of the absolute value of $G^x(r)$ along the $AB$ path for different values of $U$ at $\beta=2$. (d--f) Corresponding results along the $BC$ path. \blue{The dashed lines are guides to the exponential decay expected for short-ranged correlations. (g) Longest-distance spin correlation $G^x_{BC}(r_{\max})$ as a function of $U$. A statistically nonzero $G^x_{BC}(r_{\max})$, together with the alternating sign pattern of $G^x_{BC}(r)$, is used as a finite-size indicator for long-distance AFM correlations along the $BC$ path.} Error bars represent the standard error from 10 independent Monte Carlo Markov chain measurements.}
\label{fig:fig8}
\end{figure}

\blue{We further examine the spatial decay of the transverse spin correlation function $G^x(r)$. For a short-ranged paramagnetic state, the spin correlation is expected to decay exponentially with distance, $|G^x(r)|\sim \exp(-r/\xi)$, so that $\ln|G^x(r)|$ decreases approximately linearly with $r$ before the signal reaches the statistical noise floor. Therefore, the logarithmic plots in Figs.~\ref{fig:fig8}(a--f) are used mainly to visualize the spatial decay of the spin correlations. Figures~\ref{fig:fig8}(a--c) show $\ln|G^x(r)|$ along the $AB$ path. For all values of $\theta$ and $U$ studied here, the correlations decay rapidly with distance, and its values at the largest accessible separation $r_{\max}$ are close to zero within statistical uncertainty. This suggests that magnetic correlation along the $AB$ path remains short-ranged and no long-range magnetic order is developed. Figures~\ref{fig:fig8}(d--f) show the corresponding results along the $BC$ path. For weak compression, $\theta\geq54^\circ$, the  correlations also decay rapidly, and remain negligible at $r_{\max}$. In contrast, for stronger compression, especially for $\theta\leq52^\circ$, the magnetic correlation decays much slower and its value at large separation systematically enhanced as $U$ increases.}~\\

\blue{To quantify this behavior without relying on visual inspection of the logarithmic plots, Fig.~\ref{fig:fig8}(g) shows the correlation at the maximum accessible $BC$ distance, $G^x_{B-C}(r_{\max})$, as a function of $U$. In a short-ranged paramagnetic state, this quantity should vanish within statistical uncertainty. For $\theta\geq54^\circ$, $G^x_{B-C}(r_{\max})$ remains close to zero for all $U$. In contrast, for $\theta\leq52^\circ$, $G^x_{B-C}(r_{\max})$ becomes statistically distinguishable from zero and increases with $U$. Together with the alternating sign pattern of $G^x_{B-C}(r)$ along the $BC$ path, this provides finite-size evidence for the onset of long-range antiferromagnetic correlations. }~\\

\begin{figure}[t!]
\centering
\includegraphics[width=0.45\linewidth]{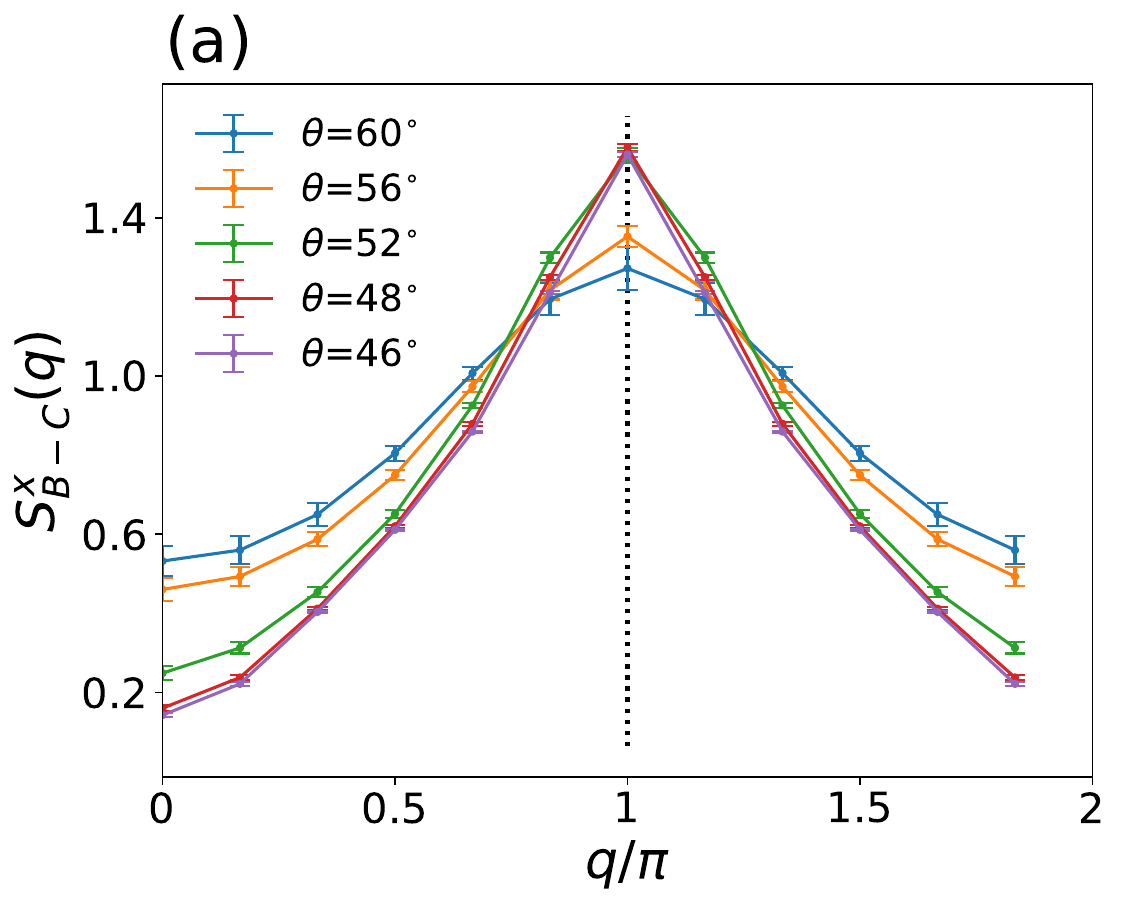}
\includegraphics[width=0.45\linewidth]{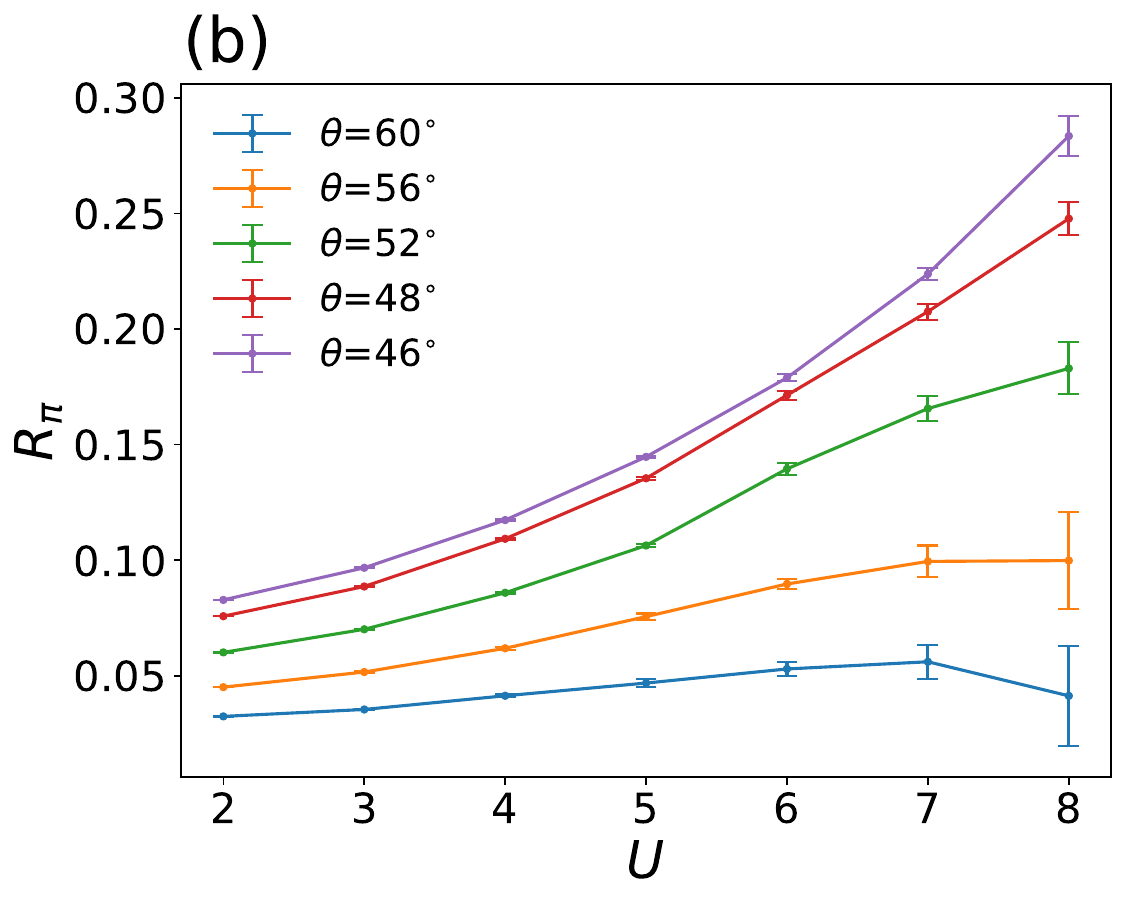}
\caption{\blue{Momentum-space characterization of the spin correlations along the BC path at $T=0.5$ and $L=6$. (a) Path-resolved static spin structure factor $S^{x}_{B-C}(q)$ at $U=7$ for different compression angles. The vertical dotted line marks the antiferromagnetic wave vector $q=\pi$. Although a maximum at $q=\pi$ is present for all shown angles, the peak becomes progressively sharper as $\theta$ decreases. (b) Symmetrized correlation ratio $R_{\pi}$ as a function of $U$ for different compression angles. The increase of $R_{\pi}$ under strong compression indicates an enhanced and sharper antiferromagnetic peak at $q=\pi$. Error bars denote the standard errors among independent Monte Carlo runs, and the lines are guides to the eye.}}
\label{fig:fig8-2}
\end{figure}

\blue{To provide an independent momentum-space characterization, we further calculate the path-resolved spin structure factor $S^{x}_{B-C}(q)$, as shown in Fig.~\ref{fig:fig8-2}(a). For all compression angles shown at $U=7$, $S^{x}_{B-C}(q)$ exhibits a maximum at $q=\pi$, reflecting the antiferromagnetic correlations already visible in real space. More importantly, the $q=\pi$ peak becomes progressively sharper as $\theta$ decreases. This change is particularly pronounced between $\theta=56^{\circ}$ and $\theta=52^{\circ}$: the peak remains relatively broad for $60^{\circ}\geq\theta\geq56^{\circ}$, whereas it becomes substantially sharper for $\theta\leq52^{\circ}$.}~\\

\blue{The evolution of the spin structure factor peak is quantified by the correlation ratio $R_{\pi}$, as shown in Fig. \ref{fig:fig8-2}(b). In the weakly compressed regime, $R_{\pi}$ remains relatively small and varies only moderately with $U$, whereas for $\theta\leq52^{\circ}$ it increases markedly as $U$ increases. The enhancement of $R_{\pi}$ occurs in the same parameter region where $G^{x}_{B-C}(r_{\max})$ becomes statistically distinguishable from zero. The real-space and momentum-space results therefore provide consistent finite-size evidence for long-range antiferromagnetic correlations.}~\\

\blue{Nevertheless, both $S^{x}_{B-C}(q)$ and $R_{\pi}$ contain contributions from short-range magnetic correlations. For example, even at $\theta=60^{\circ}$, the strengthening of the nearest-neighbor antiferromagnetic correlations with increasing $U$ produces a moderate enhancement of these quantities, although $G^{x}_{B-C}(r_{\max})$ remains consistent with zero. Therefore $S^{x}_{B-C}(q)$ and $R_{\pi}$ cannot by themselves distinguish the onset of long-range correlations from a general enhancement of short-range correlation, nor can they determine a thermodynamic critical point. We thus use them as supporting finite-size indicators in conjunction with the longest-range correlation $G^{x}_{B-C}(r_{\max})$.}~\\

\subsection{The Phase Diagram}
\label{subsec:Phase_diagram}

\begin{figure} [t!]
\includegraphics[width=0.48\linewidth]{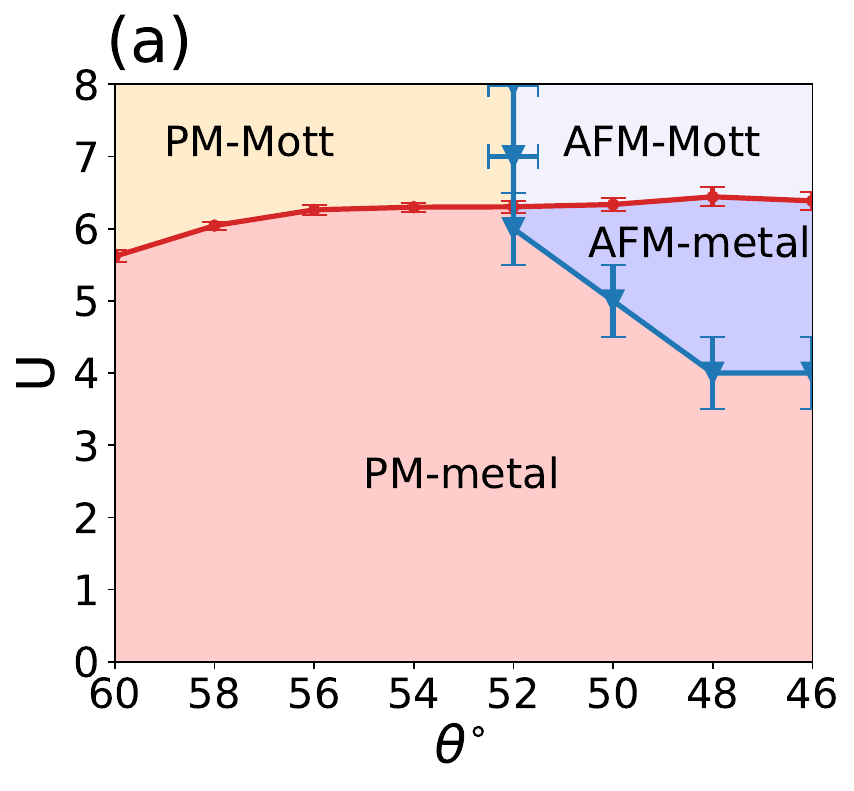}
\includegraphics[width=0.48\linewidth]{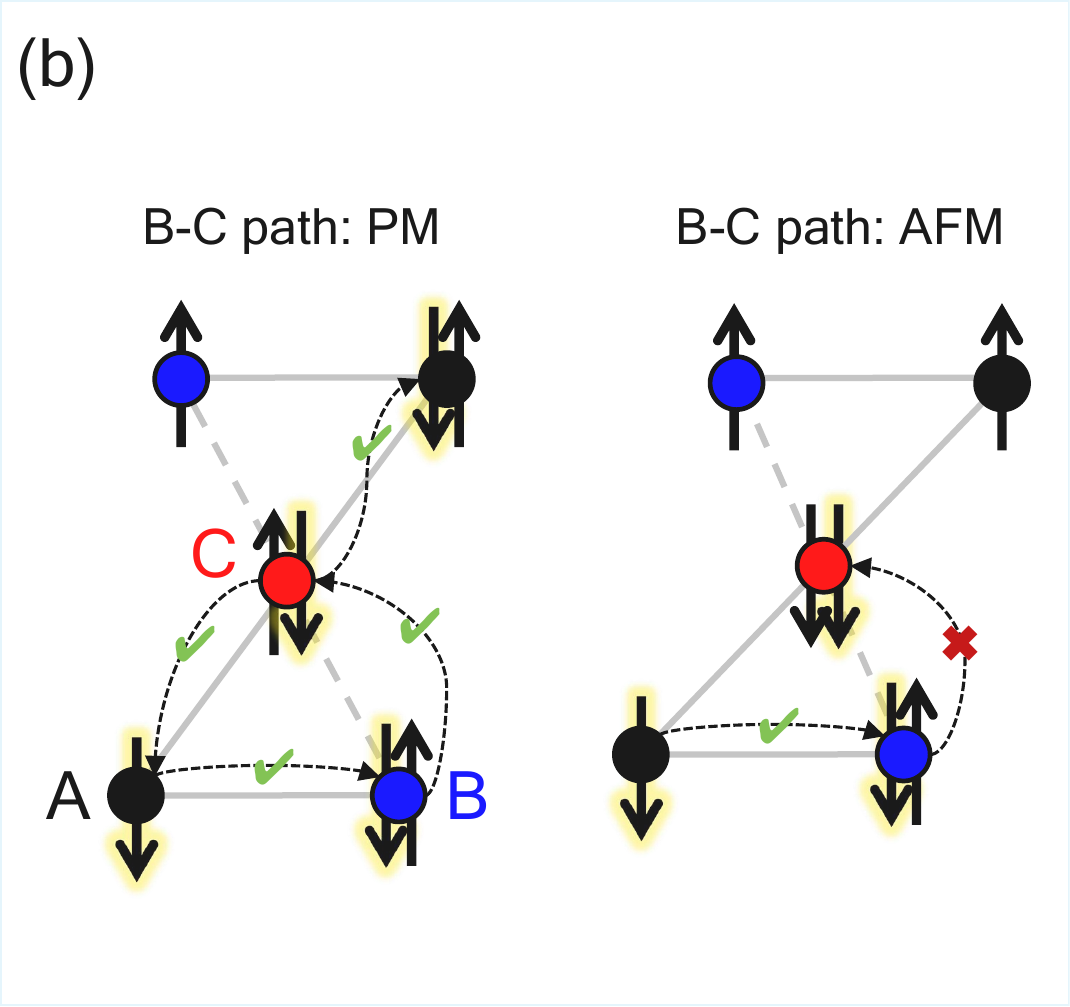}
\caption{\blue{(a) Estimated phase diagram as a function of $U$ and $\theta$. The red line marks the metal--Mott boundary inferred from the effective charge gap extracted from the low-temperature compressibility, with error bars from uncertainty propagation in the charge-gap fitting. The blue line marks the onset of long-range AFM correlations along the $BC$ path in the finite system studied, with error bars reflecting the uncertainty associated with the discrete sampling of $U$ and $\theta$. The AFM boundary is assigned primarily from the statistically nonzero $G^x_{B-C}(r_{\max})$ and the alternating sign pattern of $G^x_{B-C}(r)$.} 
(b) In the PM phase, the spin down electron highlighted in yellow can propagate between the $A$ sites through a multi-step process $A\!\to\!B\!\to\!C\!\to\!A$ (green check marks). Once a staggered AFM pattern develops along the $BC$ direction, a spin down electron arriving at sublattice $B$ is prevented from hopping to sublattice $C$ because a spin down electron on $C$ is already occupied (Pauli exclusion). As a result, the $A\!\to\!B\!\to\!C\!\to\!A$ process is strongly suppressed (red cross), which in turn reduces the conductivity of sublattice $A$.}
\label{fig:fig9}
\end{figure}

Figure~\ref{fig:fig9}(a) presents an estimated low-temperature phase diagram as a function of interaction strength $U$ and geometric parameter $\theta$ at $T=0$, which features four distinct regions : paramagnetic metal (PM–metal), paramagnetic Mott insulator (PM–Mott), antiferromagnetic metal (AFM–metal), and antiferromagnetic Mott insulator (AFM–Mott). The red solid line denotes the metal–Mott insulator phase boundary for the whole lattice, which is determined by identifying the value of $U$ where the charge gap $\Delta^c = 0$, as extracted from the temperature dependence of the electronic compressibility $\kappa(T)$ at various $\theta$. \blue{The blue line denotes the finite-size estimate of the onset of long-range AFM correlations along the $BC$ path. This boundary is assigned primarily from the growth of the longest-distance correlation $G^x_{B-C}(r_{\max})$, which becomes statistically distinguishable from zero, together with the alternating sign structure of $G^x_{B-C}(r)$. 
Because the long-range AFM correlations are most clearly resolved along the $BC$ path, we label the corresponding region as AFM in the phase diagram. This label is intended only as a finite-size characterization of the dominant magnetic correlations. A systematic finite-size analysis would be required to establish whether true long-range AFM order exists and is left for future work.}

\blue{We also comment on the order of the transitions. Since the simulations are performed at finite temperature and at a fixed finite system size ($L=6$), the present data do not allow us to unambiguously determine whether the Mott and magnetic transitions are first-order or continuous in the thermodynamic limit. Within our numerical resolution, we do not observe clear discontinuities in the double occupancy, compressibility, or spin correlation functions across the estimated boundaries. Thus, the results are consistent with a continuous transition or a smooth crossover-like evolution, although a weakly first-order transition cannot be ruled out without systematic finite-size scaling, hysteresis analysis, or Binder-ratio calculations.}~\\

The emergence of different phases originates from the competition between electronic correlation and geometric frustration. In the regime of small $U$, the kinetic energy dominates, and the system remains in a PM–metal phase: here, the upper and lower Hubbard bands overlap ($\Delta^c<0$), and the spin correlation function $\ln|G^x(r)|$ decays linearly with $r$, indicating the presence of only short-range magnetic correlations. As $U$ increases, under conditions of weak geometric frustration (small $\theta$), \blue{long-range antiferromagnetic correlation} emerges along the $BC$ path while the system remains metallic, resulting in an AFM–metal phase. Upon further increasing $U$ and crossing the Mott transition boundary (blue line), electrons become localized ($\Delta^c>0$), and the \blue{long-range antiferromagnetic correlation} persists, leading to the AFM–Mott phase. In contrast, for strong geometric frustration ($\theta$ close to $60^{\circ}$), the system remain without magnetic long-range correlation even after reaching the critical $U$ required for the Mott insulating state. \blue{Within the simulated range of $U$, we do not observe a statistically nonzero $G^x_{B-C}(r_{\max})$ for $60^{\circ}\geq\theta \geq 54^\circ$. In contrast, $G^x_{B-C}(r_{\max})$ becomes statistically distinguishable from zero on the smaller-$\theta$ side. Therefore, in the present finite-size data, the estimated AFM boundary is located near $\theta \simeq 52^\circ$.}~\\

The phase diagram reveals that, as $\theta$ varies, the Mott transition point $U^c$ of the whole lattice exhibits only small changes, whereas the paramagnetic to antiferromagnetic transition point changes significantly. However, if we consider the metal–Mott insulator transition for each sublattice individually  (Fig.~\ref{fig:fig6}(d)), we can find that their respective $U^c$ also show large variation, indicating that geometric compression of the lattice affects both conductivity and magnetism to a comparable strength. Additionally, we observe that within the simulated range of $U$, the PM–AFM transition along the $BC$ path first appears at $\theta=52^\circ$, coinciding with the $\theta$ where a significant decrease in $U^c_{A}$ for sublattice $A$ occurs.~\\

The overall evolution of the phase diagram can be discussed in terms of how the hopping network changes with $\theta$ and how increasing $U$ suppresses charge fluctuations, as indicated by the compressibility $\kappa_\alpha$ ($\alpha=A,B,C)$ and supported by the double occupancy $D_\alpha$. The sublattice transition $U^c_{\alpha}$ are defined operationally from the strong suppression of $\kappa_\alpha$ within our finite temperature data.~\\

When $\theta \ge 52^\circ$, the degree of compression is moderate and the system remains paramagnetic. The three sublattices are still strongly connected, and the global phase diagram resembles that of the standard kagome lattice. At the sublattice level, sublattice $A$ already develops a nonzero next-nearest-neighbor hopping, while sublattices $B$ and $C$ are still mainly governed by nearest-neighbor hopping (see the hopping amplitudes extracted from the geometry in Fig.~\ref{fig:fig1}). This additional hopping channel is consistent with more efficient electronic motion associated with the $A$ sublattice, so a larger $U$ is required to strongly suppress $\kappa_A$, leading to $U^c_A > U^c_{B/C}$ (see Fig.~\ref{fig:fig5} and Fig.~\ref{fig:fig6}). Meanwhile, in the same $\theta \ge 52^\circ$ regime, the low-temperature charge response on $B/C$, as captured by $\kappa_{B/C}$, is reduced and the extracted $U^c_{B/C}$ decreases (see Fig.~\ref{fig:fig5} and Fig.~\ref{fig:fig6}).~\\

When $\theta \le 52^\circ$, the compression becomes strong and the $BC$ hopping amplitude becomes much larger than all other pathways, making the $BC$ motion dominant and giving rise to spontaneous AFM order along the $BC$ direction (see Fig.~\ref{fig:fig7} and Fig.~\ref{fig:fig8}). In such a staggered spin background, the two-step hopping process $A$--$B$--$C$--$A$ is strongly suppressed at low temperature, as illustrated in the schematic diagram shown in Fig. \ref{fig:fig9}(b). Electrons starting from $A$ have difficulty propagating through the $BC$ path in the presence of staggered spins, which reduces the conductivity of sublattice $A$ and results in a rapid decrease of $U^c_A$ for $\theta \le 52^\circ$. In contrast, because the $BC$ hopping itself is strongly enhanced by compression, and the next-nearest-neighbor hopping on $B/C$ is also enhanced, sublattices $B$ and $C$ remain more conductive. As a result, a larger $U$ is needed to strongly suppress $\kappa_{B/C}$, leading to the rapid increase of $U^c_{B/C}$ when $\theta \le 52^\circ$.~\\

\blue{We finally comment on the physical scale of the distortion associated with the AFM regime. In the present geometry, the $AB$ and $AC$ bond lengths are fixed to $0.5$, while $r_{BC}(\theta)=\sin(\theta/2)$. Thus, relative to the ideal kagome lattice at $\theta=60^\circ$, the $BC$ bond is shortened by approximately $12.3\%$ at $\theta=52^\circ$ and $21.9\%$ at $\theta=46^\circ$. If this local bond shortening is roughly interpreted as a uniaxial elastic strain, a Hooke-law estimate gives a stress $\sigma\sim E\epsilon$, where $E$ is the Young's modulus~\cite{landau_elasticity,ashby2012engineering}. For a typical metal, $E\sim50$--$200$ GPa~\cite{haynes2016crc,asm1990properties}, this corresponds to stresses of order $5$--$20$ GPa for a $10\%$ strain and $10$--$40$ GPa for a $20\%$ strain. Since such large strains are likely to exceed the range of simple linear elasticity in many bulk metals, realizing the strong-compression regime by ordinary elastic strain in bulk kagome compounds would require careful materials design.}~\\

\blue{The present model should therefore be viewed as a controlled hopping-tuning model that captures the effects of enhanced hopping anisotropy and reduced geometric frustration. Experimentally, similar hopping hierarchies may be approached in several engineered settings. For example, moiré transition-metal-dichalcogenide heterobilayers can host effective kagome spin models~\cite{motruk2023kagome}, while ultracold atoms in optical kagome lattices allow direct engineering of tunneling amplitudes~\cite{jo2012ultracold}. In these systems, the relevant hopping amplitudes may be tuned either through lattice or environmental engineering or, in the case of optical lattices, by directly controlling the tunneling barriers.}

\section{Conclusion}
\label{sec:conclusion}
In this work, we investigated the half-filled Hubbard model on a diagonally compressed kagome lattice with exponentially decaying distance-dependent hopping, $t(r)=t_0e^{-r/r_0}$, using determinant quantum Monte Carlo simulations. \blue{This setting provides a controlled way to tune the hierarchy of hopping amplitudes through the lattice angle $\theta$.} By analyzing sublattice-resolved double occupancy, electronic compressibility, and the low-temperature behavior of $\kappa(T)$, we find that the compressed hopping geometry can induce a pronounced sublattice-selective Mott response.~\\

\blue{For $\theta$ close to $60^\circ$, the three sublattices remain nearly equivalent. As the lattice is compressed, the hopping network becomes sublattice dependent. In the weakly compressed regime, the $A$ sublattice develops additional long-range hopping channels earlier than the $B$ and $C$ sublattices, leading to a larger critical interaction $U^c_A$ compared with $U^c_{B/C}$. In the strongly compressed regime, $\theta\lesssim52^\circ$, the $BC$ path develops long-range AFM correlations in the finite-size spin-correlation data. These correlations suppress the effective $A\!\to\!B\!\to\!C\!\to\!A$ hopping process and reduce the metallicity of the $A$ sublattice, leading to a rapid decrease of $U^c_A$. In contrast, the enhanced $BC$ hopping keeps the $B/C$ sublattices more conductive, so that a larger interaction is required to suppress their compressibility.}~\\

We further investigated the magnetic properties through the transverse spin correlation function. Along the $AB$ path, $G^x(r)$ decays rapidly for all parameters studied, consistent with short-ranged paramagnetic behavior along this direction. \blue{Along the $BC$ path, the correlations also remain short-ranged for weak compression. However, for stronger compression and sufficiently large $U$, the longest-distance correlation $G^x_{B-C}(r_{\max})$ becomes statistically distinguishable from zero and increases with $U$, while $G^x_{B-C}(r)$ shows an alternating sign pattern. We interpret this behavior as evidence for the onset of long-range AFM correlations along the $BC$ path in the finite system studied. 
The sharpening of the $q=\pi$ peak in $S^{x}_{B-C}(q)$ and the enhancement of $R_{\pi}$ provide a consistent momentum-space of the AFM signature.}

\blue{Combining the charge and spin responses, we construct an estimated $U$--$\theta$ phase diagram. The metal--Mott insulator boundary is inferred from the charge gap extracted from the compressibility, while the AFM boundary is estimated from the onset of statistically resolvable long-range AFM correlations along the $BC$ path. The resulting PM-metal, PM-Mott, AFM-metal, and AFM-Mott labels distinguish regimes with different charge responses and dominant magnetic correlations. Establishing the thermodynamic nature of these regimes and their boundaries will require systematic finite-size scaling analysis and lower-temperature simulations., which we leave for future work.}~\\

\blue{Finally, the strongly compressed regime should be interpreted primarily as a controlled hopping-anisotropy limit rather than a realistic elastic-strain realization of bulk kagome metals, since such a large modulation of lattice geometry and hopping amplitudes would be difficult to achieve through ordinary elastic strain. It may instead be more naturally accessible in moir'e or heterostructure platforms, or in ultracold-atom kagome simulators with engineered tunneling amplitudes.}~\\

\begin{acknowledgments}
We acknowledge financial support from Research Grants Council of Hong Kong (Grant No. CityU 11318722), National Natural Science Foundation of China (Grant No. 12005179, 12204130), City University of Hong Kong (Grant No. 9610438, 7005610, 7020156), the Shenzhen Fundamental Research Program (No.~JCYJ20250604145655074) and Shenzhen Key Laboratory of Advanced Functional Carbon Materials Research and Comprehensive Application~(No.~ZDSYS20220527171407017).

\end{acknowledgments}

\nocite{*}



%

\clearpage

\appendix

\section{\blue{Fitting of the compressibility}}
\label{sec:kappa_fit}

\begin{figure} [t!]
\includegraphics[width=0.32\linewidth]{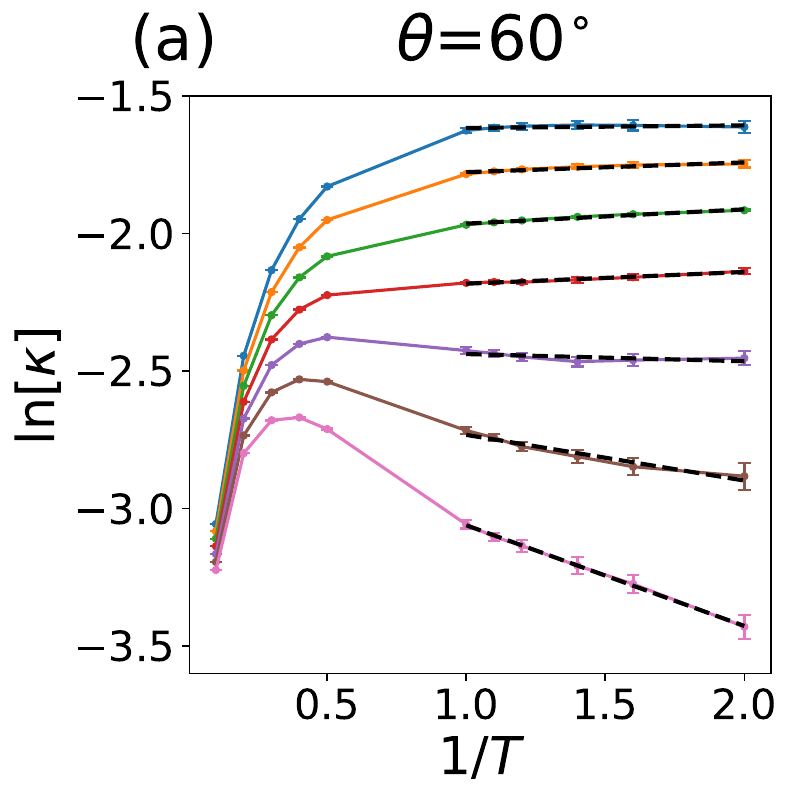}
\includegraphics[width=0.32\linewidth]{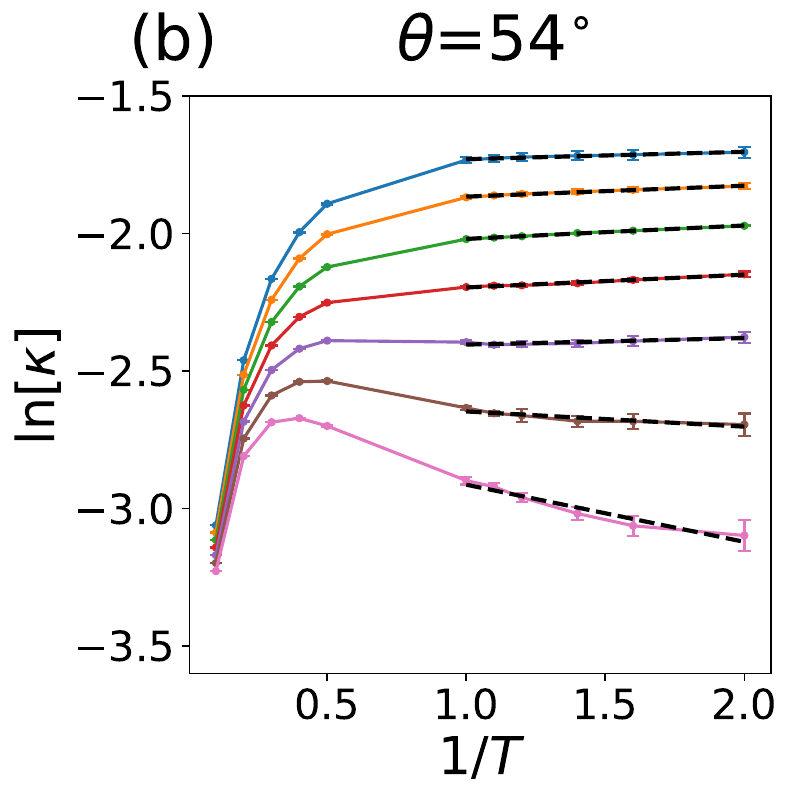}
\includegraphics[width=0.32\linewidth]{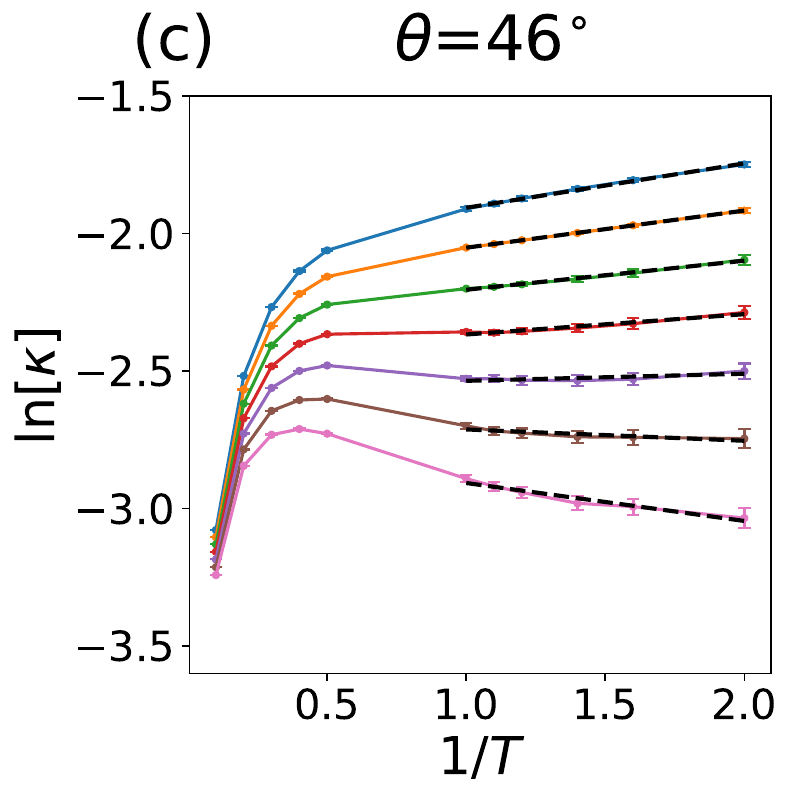}
\caption{\blue{Representative fits of the lattice-averaged compressibility. The data in Fig.~5 are replotted as \(\ln\kappa\) versus \(1/T\) for (a) \(\theta=60^\circ\), (b) \(\theta=54^\circ\), and (c) \(\theta=46^\circ\). The dashed black lines denote linear fits in the low-temperature regime \(T<1\), based on \(\ln\kappa=\ln k_1-\Delta^c/T\). The fitted slope is \(-\Delta^c\). A negative slope therefore corresponds to a positive effective charge gap, for which \(\kappa(T)\) vanishes exponentially in the \(T\to0\) limit.}}
\label{fig:appendix.A}
\end{figure}

\blue{In the main text, the metal-Mott insulator boundary is determined from the effective charge gap $\Delta^{c}$ extracted from the low-temperature behavior of the electronic compressibility. Here we provide representative fits to clarify the procedure. We note that, in a finite-size finite-temperature DQMC calculation, the compressibility \(\kappa(T)\) is not expected to vanish at any finite temperature. Therefore, the insulating behavior is not inferred from the absolute magnitude of \(\kappa\) at finite \(T\). Instead, we examine whether the low-temperature behavior is consistent with the functional form,
\begin{equation}
    \kappa(T)=k_1 \exp(-\Delta^c/T),
\end{equation}
where \(\Delta^c\) is the effective charge gap. Taking the logarithm gives
\begin{equation}
    \ln \kappa = \ln k_1 - \frac{\Delta^c }{T}.
\end{equation}
Thus, in a plot of \(\ln \kappa\) versus \(1/T\), the fitted slope in the low-temperature regime is \(-\Delta^c\).}

\blue{
Figure~\ref{fig:appendix.A} shows representative fits for the lattice-averaged compressibility at \(\theta=60^\circ\), \(54^\circ\), and \(46^\circ\). The dashed black lines are linear fits performed in the low-temperature regime \(T<1\), using the same fitting procedure as that used to obtain the effective charge gaps shown in Figure~\ref{fig:fig6}. A negative slope corresponds to a positive \(\Delta^c\), for which \(\kappa(T)\) is exponentially suppressed as \(T\to0\). In contrast, a non-negative fitted slope indicates that no positive activation gap is resolved within this fitting procedure, corresponding to a compressible metallic regime.}

\section{Comparison between the correlation functions $G^x(r)$ and $G^z(r)$}
\label{sec:sz_sx}

\begin{figure} [t!]
\includegraphics[width=0.49\linewidth]{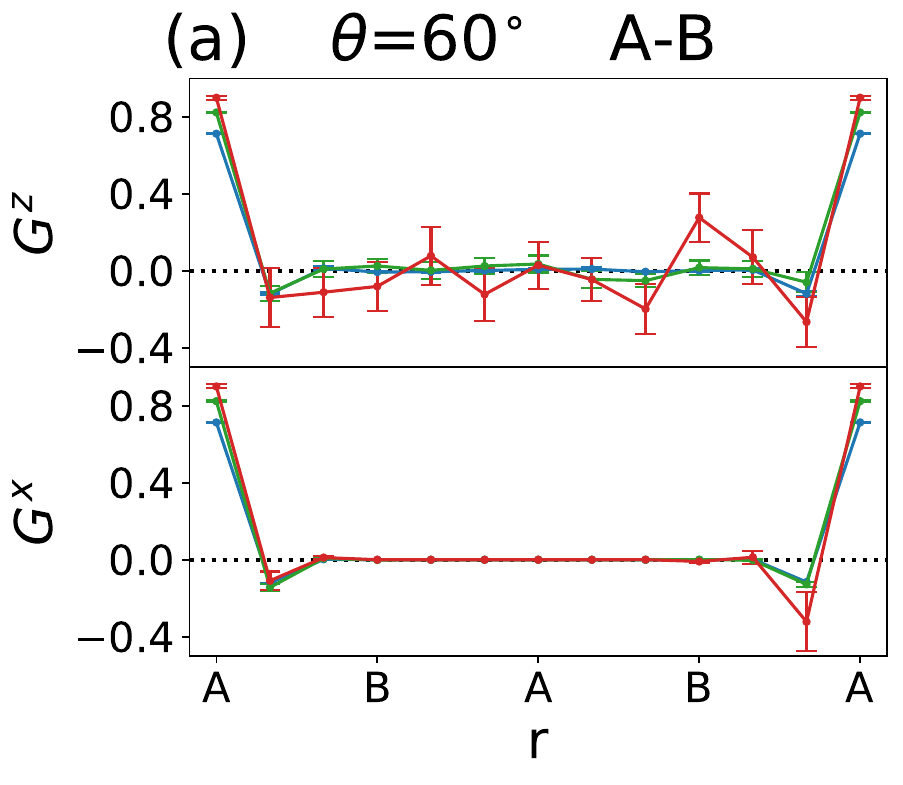}
\includegraphics[width=0.49\linewidth]{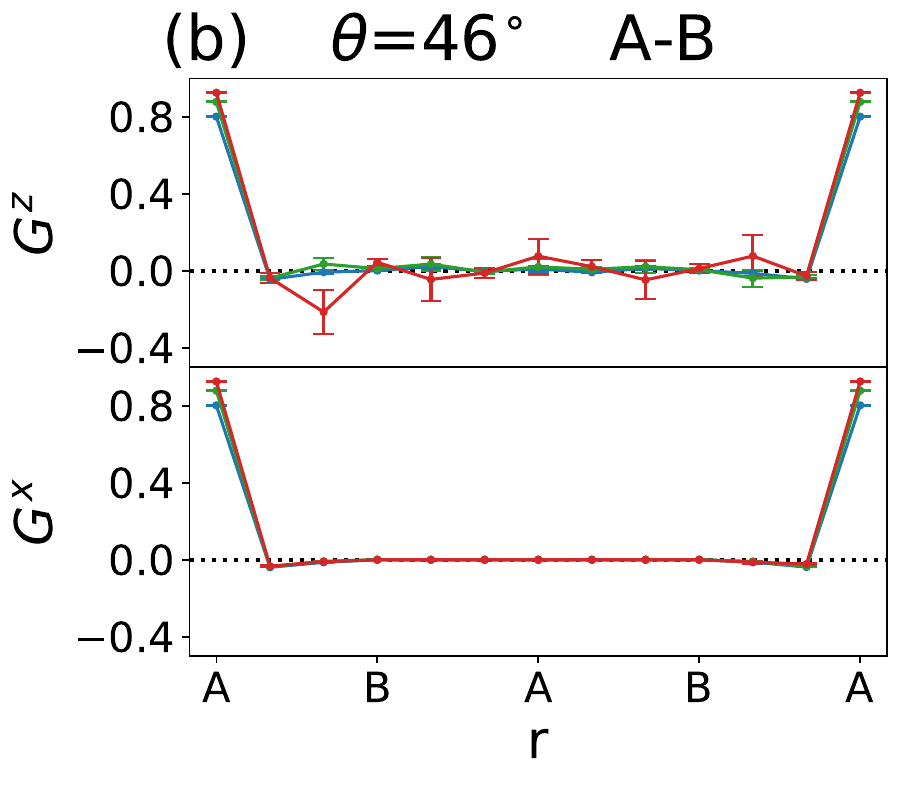}
\includegraphics[width=0.49\linewidth]{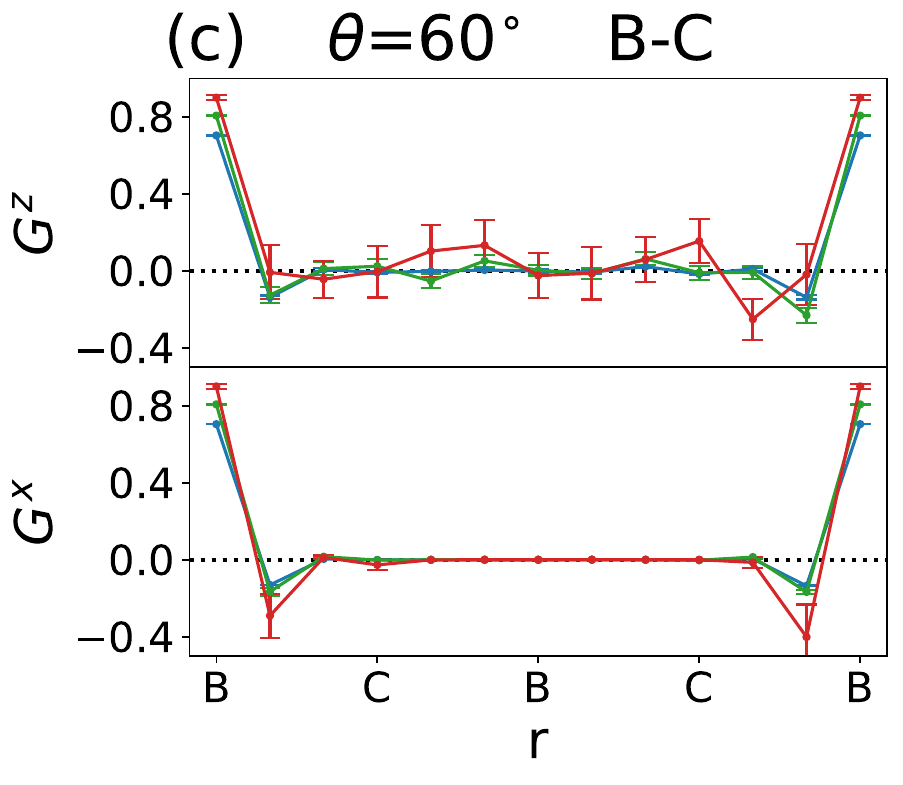}
\includegraphics[width=0.49\linewidth]{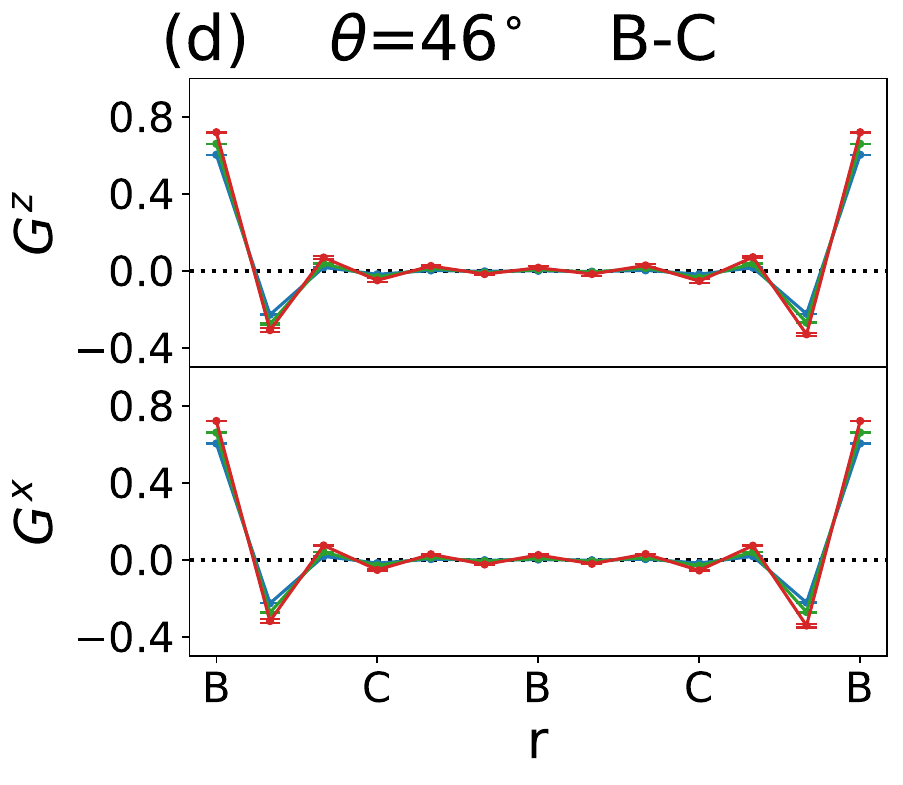}
\includegraphics[width=0.493\linewidth]{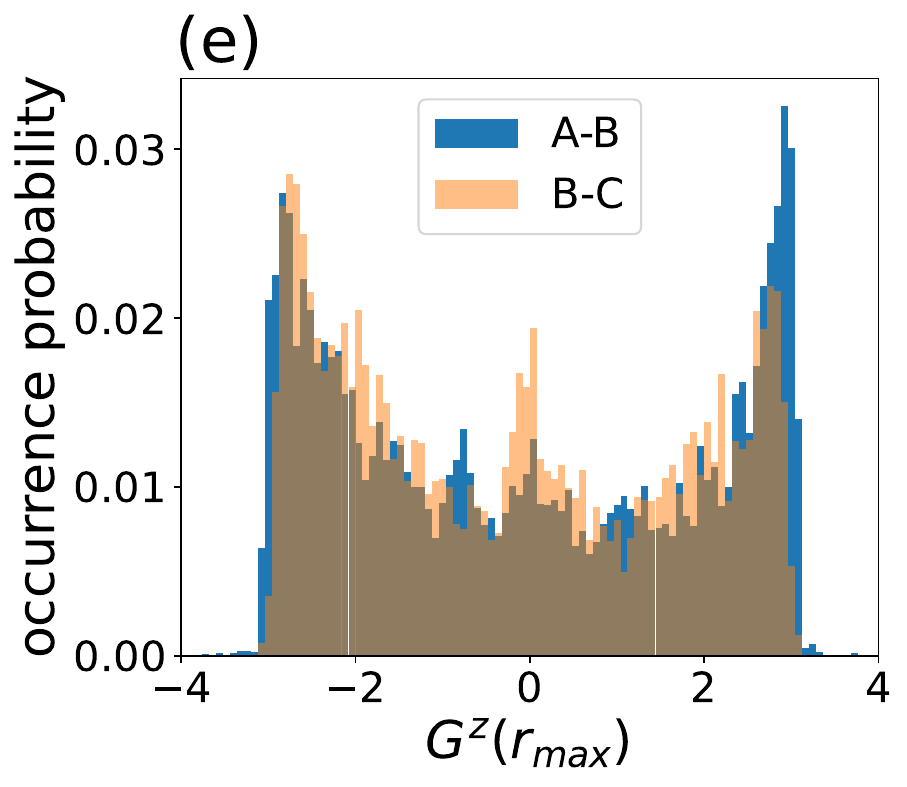}
\includegraphics[width=0.487\linewidth]{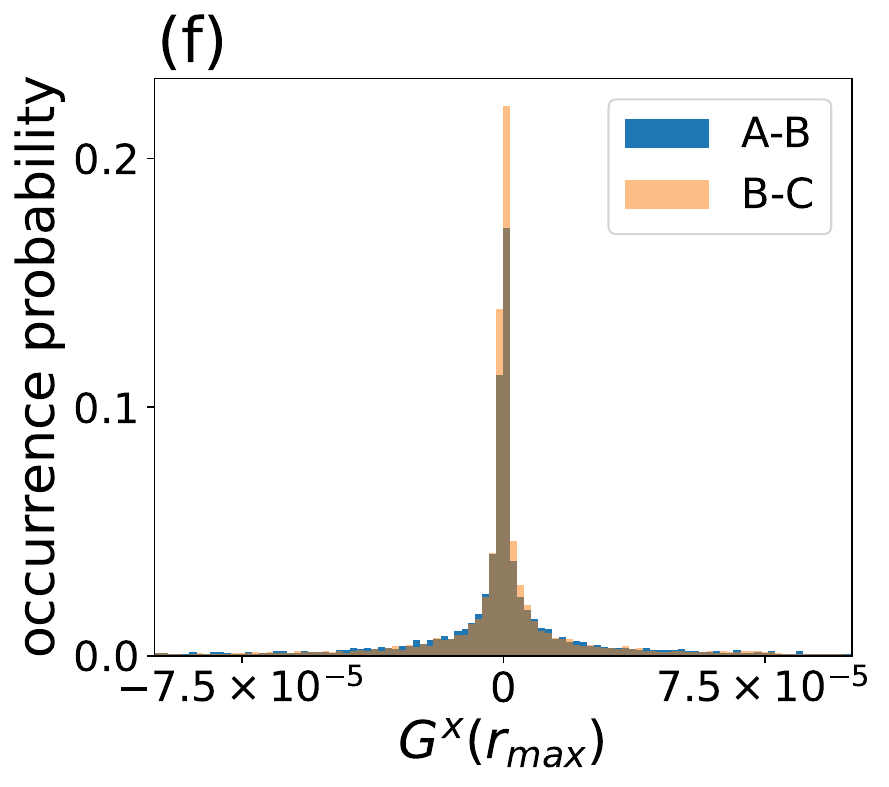}
\caption{Panels (a) and (b) show the spin correlation functions along the $A$–$B$ path for $\theta=60^\circ$ and $\theta=46^\circ$, respectively. The upper panel depicts the longitudinal correlation $G^{z}(r)$, and the lower panel presents the transverse correlation $G^{x}(r)$, with line colors denoting different values of $U$ as in Fig.~\ref{fig:fig7}(a). Panels (c) and (d) display the corresponding data for the $B$–$C$ path. Panels (e) and (f) show the DQMC sampling distributions of $G^{z}(r_{\max})$ and $G^{x}(r_{\max})$ at the largest distance $r_{\max}$ for $\theta=60^\circ$ and $U=8$.}
\label{fig:appendix.B}
\end{figure}

Figures~\ref{fig:appendix.A}(a–d) compare the distance dependence of $G^{z}(r)$ and $G^{x}(r)$ for different angles and paths. As shown in Figs.~\ref{fig:appendix.A}(a–c), in the absence of long-range magnetic order, $G^{z}$ exhibits measurement values and uncertainties comparable to $G^{x}$ only at $r=0$ and $r=r_{AB}$ (or $r=r_{BC}$), where $r_{AB}$ (or $r_{BC}$) denotes the distance between earest neighboring sublattices. As $r$ increases, $G^{z}$ fluctuates strongly around zero, and for $U=8$, the error bar is typically around 0.2, rendering it difficult to resolve long-range order. Only for the $B$–$C$ path at $\theta=46^\circ$, where antiferromagnetic order emerges, do $G^{z}$ and $G^{x}$ display similar measurement values and uncertainties at all distances. Overall, using $G^{x}$ provides a more robust and reliable characterization of magnetic behavior in the system.~\\

From the perspective of sampling, the difference in the uncertainty between $G^{z}$ and $G^{x}$ primarily arises from their different sampling ranges. Figures~\ref{fig:appendix.A}(e) and (f) show the sampling distributions of $G^{z}$ and $G^{x}$ at the maximal distance $r_{\max}$ for $\theta=60^\circ$ and $U=8$. It is evident that, for both the $A$–$B$ and $B$–$C$ paths, the $G^{z}$ distribution spans approximately $[-3, 3]$, with high-probability regions near $\pm3$, while $G^{x}$ exhibits a sharp peak at zero with a width of only $O(10^{-5})$. Achieving a $G^{z}$ value with statistical accuracy comparable to $G^{x}$ (i.e., $G^{z} \approx 0$) would require highly symmetric sampling about zero over the interval $[-3,3]$, which clearly necessitates a vastly greater number of samples. Therefore, the statistical uncertainty in $G^{z}$ is much larger than in $G^{x}$.~\\

From a definitional perspective, $G^{x}$ contains only off-diagonal terms of the Green’s function (see Eq.~\eqref{eq:Gx_r}), which vary little among different $h$. In contrast, $G^{z}$ contains not only off-diagonal terms but also the diagonal term $\langle S_i^z \rangle \langle S_{i+r}^z \rangle$ (see Eq.~\eqref{eq:Gz_r}), whose value fluctuates strongly at half-filling, leading to much larger sampling noise in $G^{z}$ compared to $G^{x}$.~\\

\end{document}